\documentclass[twocolumn]{aastex631}

\newcommand{\eg}{e.g.}

\received{2024 June 4}
\revised{2024 December 3}
\accepted{2025 January 3}
\published{2025 February 6}
\submitjournal{ApJ}
\shorttitle{Cluster candidates at $z\sim2$}
\shortauthors{Kiyota et al.}
\graphicspath{{./}{figures/}}

\begin{document}

\title{Cluster candidates with massive quiescent galaxies at $z\sim2$}
\correspondingauthor{Tomokazu Kiyota}
\email{tomokazu.kiyota@grad.nao.ac.jp}

\author[0009-0004-4332-9225]{Tomokazu Kiyota}
\affiliation{Department of Astronomical Science, The Graduate University for Advanced Studies, SOKENDAI, 2-21-1 Osawa, Mitaka, Tokyo, 181-8588, Japan}
\affiliation{National Astronomical Observatory of Japan, 2-21-1 Osawa, Mitaka, Tokyo, 181-8588, Japan}

\author[0000-0002-4225-4477]{Makoto Ando}
\affiliation{National Astronomical Observatory of Japan, 2-21-1 Osawa, Mitaka, Tokyo, 181-8588, Japan}

\author[0000-0002-5011-5178]{Masayuki Tanaka}
\affiliation{National Astronomical Observatory of Japan, 2-21-1 Osawa, Mitaka, Tokyo, 181-8588, Japan}
\affiliation{Department of Astronomical Science, The Graduate University for Advanced Studies, SOKENDAI, 2-21-1 Osawa, Mitaka, Tokyo, 181-8588, Japan}

\author[0000-0002-4606-5403]{Alexis Finoguenov}
\affiliation{Department of Physics, University of Helsinki, P.O. Box 64, FI-00014 Helsinki, Finland}

\author[0000-0003-3883-6500]{Sadman Shariar Ali}
\affiliation{Subaru Telescope,  National Astronomical Observatory of Japan, 650 North Aohoku Place, Hilo, HI, 96720, USA}

\author{Jean Coupon}
\affiliation{Department of Astronomy, University of Geneva, ch. d’Écogia 16, CH-1290 Versoix, Switzerland}

\author[0000-0001-8325-1742]{Guillaume Desprez}
\affiliation{Department of Astronomy and Astrophysics and Institute for Computational Astrophysics, Saint Mary's University, 923 Robie Street, Halifax, NS B3H 3C3, Canada}

\author[0000-0001-8221-8406]{Stephen Gwyn}
\affiliation{NRC Herzberg Astronomy and Astrophysics, 5071 West Saanich Road, Victoria, BC V9E 2E7, Canada}

\author[0000-0002-7712-7857]{Marcin Sawicki}
\affiliation{Department of Astronomy and Astrophysics and Institute for Computational Astrophysics, Saint Mary's University, 923 Robie Street, Halifax, NS B3H 3C3, Canada}

\author[0000-0003-4442-2750]{Rhythm Shimakawa}
\affiliation{Waseda Institute for Advanced Study (WIAS), Waseda University, 1-21-1 Nishi-Waseda, Shinjuku, Tokyo, 169-0051, Japan}
\affiliation{Center for Data Science, Waseda University, 1-6-1 Nishi-Waseda, Shinjuku, Tokyo, 169-0051, Japan}

\begin{abstract}
Galaxy clusters are crucial to understanding the role of the environment in galaxy evolution. However, due to their rarity, only a limited number of clusters have been identified at $z\gtrsim2$. In this paper, we report a discovery of seven cluster candidates with massive quiescent galaxies at $z\sim2$ in the $3.5\,\mathrm{deg}^{2}$ area of the XMM-LSS field, roughly doubling the known cluster sample at this frontier redshift if confirmed. We construct a photometric redshift catalog based on deep ($i\sim26$, $K_\mathrm{s}\sim24$) multiwavelength photometry from $u^*$-band to $K$-band gathered from the Hyper Suprime-Cam Subaru Strategic Program and other collaborative/public surveys. We adopt a Gaussian kernel density estimate with two different spatial scales ($10\arcsec$ and $60\arcsec$) to draw a density map of massive ($\log(M_{*}/M_{\odot})>10.5$) and quiescent ($\log(\mathrm{sSFR\,[\mathrm{yr^{-1}}]})<-10$) galaxies at $z\sim2$. Then, we identify seven prominent overdensities. These candidates show clear red sequences in color-magnitude diagrams ($z-H$ versus $H$). Moreover, one of them shows an extended X-ray emission with $L_\mathrm{X}=(1.46\pm0.35)\times10^{44}$~erg~s$^{-1}$, suggesting its virialized nature. There is no clear evidence of enhancement nor suppression of the star formation rate of the main sequence galaxies in the clusters. We find that cluster galaxies have a higher fraction of transition population with $-10.5<\log(\mathrm{sSFR\,[\mathrm{yr^{-1}}]})<-10$ ($12\%$) than the field ($2\%$), which implies the ongoing star formation quenching. The quiescent fraction in the cluster candidates also exceeds that in the field. We confirm that the excess of a quiescent fraction is larger for higher-mass galaxies. This is the first statistical evidence for the mass-dependent environmental quenching at work in clusters even at $z\sim2$. 
\end{abstract}


\section{Introduction} \label{sec:intro}

Galaxies have large diversities in their luminosity, color, morphology, etc. Some galaxies are represented by red colors, quenched star formation, and elliptical shapes, and some are characterized by blue colors, active star formation, and spiral morphologies. In the local Universe, it is well known that red galaxies mainly reside in galaxy clusters, while blue galaxies are more likely to be located in a field environment (\eg, \citealt{dressler80, butcher84, goto03}; and \citealt{blanton09} for a review). This morphology-density, color-density, or even star formation-density relation suggests that galaxies are affected by the surrounding environment during their formation and evolution (\eg, \citealt{peng10,peng12,wetzel12}). These relations have also been investigated beyond the local Universe (\eg, \citealt{dressler97, kauffmann04, smith05, cooper06, fossati17, kawinwanichakij17, chartab20, sazonova20, mei23}). This segregation of galaxies is likely due to several different physical processes: hierarchical structure formation, interaction between galaxies and intracluster medium (ICM), galaxy mergers, and active galactic nuclei (AGN) activities (\eg, \citealt{croton06, gunn72, boselli06, moore96, moore98, fabian12, bahe15}). However, how and when environment plays a role in the evolution of galaxies is not fully understood yet.

It has been known for some time that the cosmic star formation density peaks at $z=1$--$3$ (\citealt{sawicki97, giavalisco04}; and see \citealt{madau14} for a review), dubbed the cosmic noon. Similarly, the connection between star formation and environment may have changed through the cosmic age. Some studies have reported the `reversal' of star formation-density relation at $z\gtrsim1$ (\eg, \citealp{elbaz07,hwang19,lemaux22,shi24}). Indeed, active star formation rather than suppression has been observed in overdense regions of galaxies at $z\gtrsim 2.5$ (i.e. protoclusters, \eg, \citealp{wang16,shimakawa18,oteo18,miller18,ito20,toshikawa24}). On the other hand, there are opposite studies that have shown an enhanced fraction of quiescent galaxies in overdense regions at these redshifts (\eg, \citealp{kodama07,newman14,cooke16,muldrew18,ando20,ando2022,ito23a,tanaka23}). The cosmic noon might be a transition epoch of star formation enhancement/suppression in dense environments, and there may be diversities in the effects of environment. To shed light on the origins of environmental effects in the densest environment in the Universe, it is important to examine galaxy properties in various clusters at this epoch statistically.

Until now, many surveys have attempted to search for clusters beyond $z=1$. For instance, the Spitzer Adaptation of the Red-sequence Cluster Survey (SpARCS; \citealt{muzzin09, wilson09}) has used red sequence galaxies as tracers of $z>1$ clusters. A large number of cluster candidates at  $z>1$ are identified through the Sunyaev–Zel'dovich effect \citep{sunyaev72} by the South Pole Telescope (SPT) cluster survey (\eg, \citealp{brodwin10,foley11,bayliss16}). Extended X-ray emissions are also used as signposts of high-redshift clusters/groups (\eg, \citealt{pierre16,pacaud16,gozaliasl19}), although the sample is limited to $z\lesssim1.5$ due to sensitivity. Large spectroscopic follow-up campaigns for cluster/group candidates at $1.0<z<1.5$, mainly composed of the SpARCS and SPT sample, have been conducted by the Gemini Cluster Astrophysics Spectroscopic Survey (GCLASS; \citealt{muzzin12}) and the Gemini Observations of Galaxies in Rich Early Environments (GOGREEN) survey (\citealt{balogh17, balogh21}). They have confirmed more than ten clusters with about one thousand confirmed members in total. 

Thanks to the growing number of distant cluster samples, the properties of cluster galaxies up to $z\sim1.5$ have been investigated statistically. The stellar mass function in clusters has a high-mass excess from that in the field (\eg, \citealt{vanderburg13,vanderburg20}). The fraction of quiescent galaxies in clusters exceeds that in the field, and the excess depends on halo mass at $1<z<1.5$ \citep{reeves21}. Interestingly, environmental quenching is more effective for higher-mass galaxies at $1<z<1.5$, which is inconsistent with a picture of environmental quenching in low redshift ($z<1$) clusters, where the environment works independently to stellar mass \citep{peng10, vanderburg18}. In these clusters, even star-forming galaxies are likely to have suppressed star formation compared to the field \citep{old20}, though not conclusive (\citealt{nantais20}). These pieces of evidence thus suggest that significant environmental quenching occurs in clusters up to $z\sim1.5$. However, higher redshift clusters remain largely unexplored. 

The redshift records of confirmed clusters are at $z\sim2$. For example, \citet{newman14} have spectroscopically confirmed the rich cluster (JKCS 041) at $z=1.80$. This cluster is also detected with extended X-ray emission by the Chandra X-ray Observatory (Chandra; \citealt{weisskopf02}) and hosts 19 confirmed member galaxies, of which 15 are quiescent galaxies (see also \eg, \citealt{andreon09, andreon14}). \citet{willis20} have reported a similar cluster at $z=1.98$ (XLSSC 122) with ICM detection through the Sunyaev-Zel'dovich effect \citep{mantz18}. The halo mass of this cluster has been estimated to be $M_\mathrm{halo}\sim10^{15}~M_\odot$ at $z=0$ \citep{willis20}; comparable to the mass of the Coma Cluster. In addition, there are other reports of mature clusters at $z\sim2$: C1 J1449+0856 at $z=2.0$ (\citealt{gobat11, gobat13, strazzullo13}); IDCS J1426.5+3508 at $z=1.8$ (\citealt{stanford12}), IDCS J1433.2+3306 at $z=1.9$ (\citealt{zeimann12}), a cluster in the COSMOS field at $z=2.1$ (\citealt{spitler12, yuan14}), Spiderweb protocluster at $z=2.2$ (\citealt{tozzi22, mascolo23}). Despite these efforts, there is only a too small sample to discuss environmental effects statistically. A larger cluster sample is needed to fully understand the role of environmental effects at this frontier redshift. 

In this paper, we report the discovery of seven cluster candidates at $z\sim 2$ with massive quiescent galaxies in the $\sim3.5\,\mathrm{deg}^{2}$ XMM-LSS field. We first construct a photometric redshift catalog based on wide and deep multiwavelength photometric data covering from optical to near-infrared. We then assume red sequence galaxies well trace evolved clusters and search for overdensities of massive quiescent galaxies, identifying seven prominent candidates at $z\sim2$. Interestingly, extended X-ray emission is detected from one of them. They are good test beds to examine the role of environmental quenching and its physical origins at this frontier redshift of $z\sim2$ 
\footnote{The terms `cluster' and `protocluster' are frequently used to distinguish between gravitationally bound and unbound systems at high redshift. While further observations are necessary to confirm whether the overdensities identified in this study are gravitationally bound, we refer to our samples as ‘cluster candidates' since they represent significant overdensities on the cluster-halo scale, approximately 1 Mpc.}

The structure of this paper is as follows. Section~\ref{sec:data} presents the data and photometric redshift catalog construction. Our cluster-finding method and identified cluster candidates at $z\sim2$ are described in Section~\ref{sec:analysis}. In this section, we also report the detection of extended X-ray emission from one of the candidates. In Section~\ref{sec:discussion}, we discuss the properties of galaxies associated with these candidates and the environmental effects at $z\sim2$. Section~\ref{sec:conclusions} summarizes and concludes the paper. Throughout this paper, we use \citet{chabrier03} initial mass function and a flat $\Lambda$CDM cosmology with $H_0 = 70 \mathrm{~km ~s^{-1} ~Mpc^{-1}}$, $\Omega_m = 0.3$ and $\Omega_\Lambda = 0.7$. All magnitudes are in the AB system \citep{oke83}. 

\section{Data} \label{sec:data}
\subsection{Photometry}
We base our analysis on a wide and deep multiwavelength catalog from the $U$-band to the $K$-band
in the Deep and UltraDeep (D/UD) fields of the Hyper Suprime-Cam Subaru Strategic Program (HSC-SSP; \citealt{aihara18a,aihara18b,aihara19,aihara22}). One of the collaborating surveys of HSC-SSP observed the D/UD fields in the $U$-band using CFHT Megacam (CLAUDS; \citealt{sawicki19})\footnote{\url{https://www.clauds.net/}} and another survey covered the same fields in the near-IR using UKIRT WFCAM (DUNES$^2$; Egami et al. in prep.).  We have compiled data from all these surveys as well as public surveys such as UKIDSS-DXS/UDS DR10 \citep{hewett06,lawrence07,casali07,hambly08}, VISTA-VIDEO DR4 \citep{jarvis13}, and UltraVISTA DR4 \citep{mccracken12}, to construct a multiwavelength catalog of the D/UD fields. We use the HSC-SSP images from the public data release 2 (PDR2; \citealt{aihara19}). 
The details of the catalog are described elsewhere (Suzuki et al. in prep.). In short, HSC images are processed with hscPipe v6 \citep{bosch18,aihara19}, while others are processed with the pipelines for each instrument. All the coadd images are first registered to the same pixel coordinates (HSC's tracts and patches).
We then construct model PSFs from each coadd image using bright stars, add the variance and pixel mask images, and run hscPipe v6 to detect and measure sources  (see \citealt{desprez23} for details).

\begin{deluxetable}{lcc}
    \tablecaption{Summary of the data set \label{tab:data}}
    \tablehead{
    \colhead{Instrument} & \colhead{Filter} & \colhead{5$\sigma$ depth} \\
    \colhead{(Survey name)} & \colhead{} & \colhead{[mag]}
    }
    \startdata
    CFHT/MegaCam & $u^{*}$ & 27.2 \\
    (CLAUDS)     &  &  \\
    \hline
    Subaru/HSC & $g$ & 26.8 \\
    (HSC-SSP)  & $r$ & 26.2 \\
               & $i$ & 25.8 \\
               & $z$ & 25.6 \\
               & $y$ & 24.5 \\
    \hline
    VISTA/VIRCAM & $Y$ & 25.2 \\
    (VIDEO)      & $J$ & 24.9 \\
                 & $H$ & 24.3 \\
                 & $K_\mathrm{s}$ & 24.1 \\
    \enddata
    \tablecomments{The $5\sigma$ limiting magnitudes are measured in blank sky with $1.5\arcsec$ diameter aperture.}
\end{deluxetable}

\subsection{Photometric redshift and stellar mass completeness}
 
We run a Bayesian photometric redshift code from \citet{tanaka15} to compute photometric redshifts as well as to infer physical properties of galaxies such as stellar mass and star formation rate (SFR) in a self-consistent manner. We use model templates from \citet{bruzual03} assuming the \citet{calzetti94} dust attenuation curve. Emission lines are added to the templates assuming solar metallicity. We calibrate our photometric redshifts (photo-$z$'s) against the many-band, high-accuracy photo-$z$'s from the Cosmic Evolution Survey (COSMOS2020; \citealp{scoville07,weaver20,ito22}) as there is no sufficiently deep spectroscopic redshift (spec-$z$) catalog.

This calibration is performed in COSMOS, but there may be a field-field variation of the photometric zero points, which degrades our photo-$z$ performance in the other D/UD fields. We estimate the photometric zero points in each field using spec-$z$'s from the Sloan Digital Sky Survey (SDSS; \citealt{york00,strauss02}) by running our code with redshifts fixed to spec-$z$'s and evaluating the residuals between models and observed photometry. We find that the variation is small (typically $<0.05$~mag), but we apply these offsets to all fields.

\begin{figure*}
\centering
\includegraphics[width=80mm]{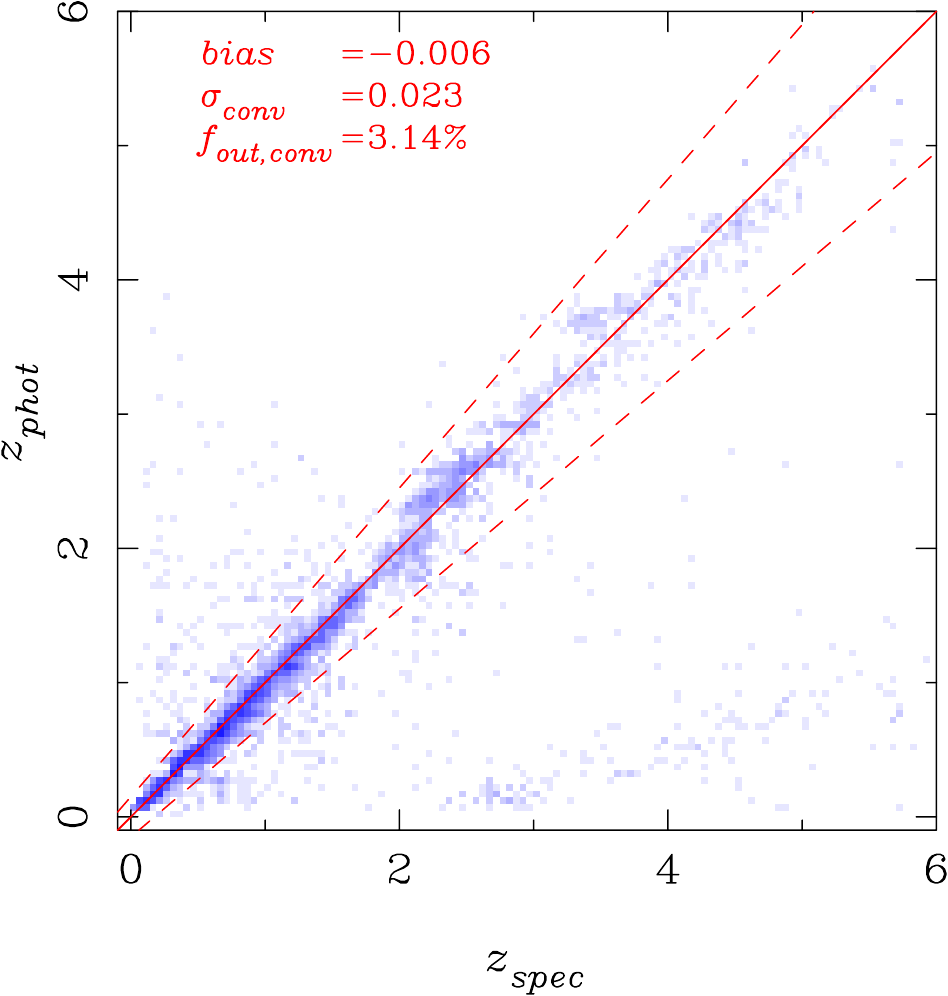}\hspace{0.5cm}
\includegraphics[width=80mm]{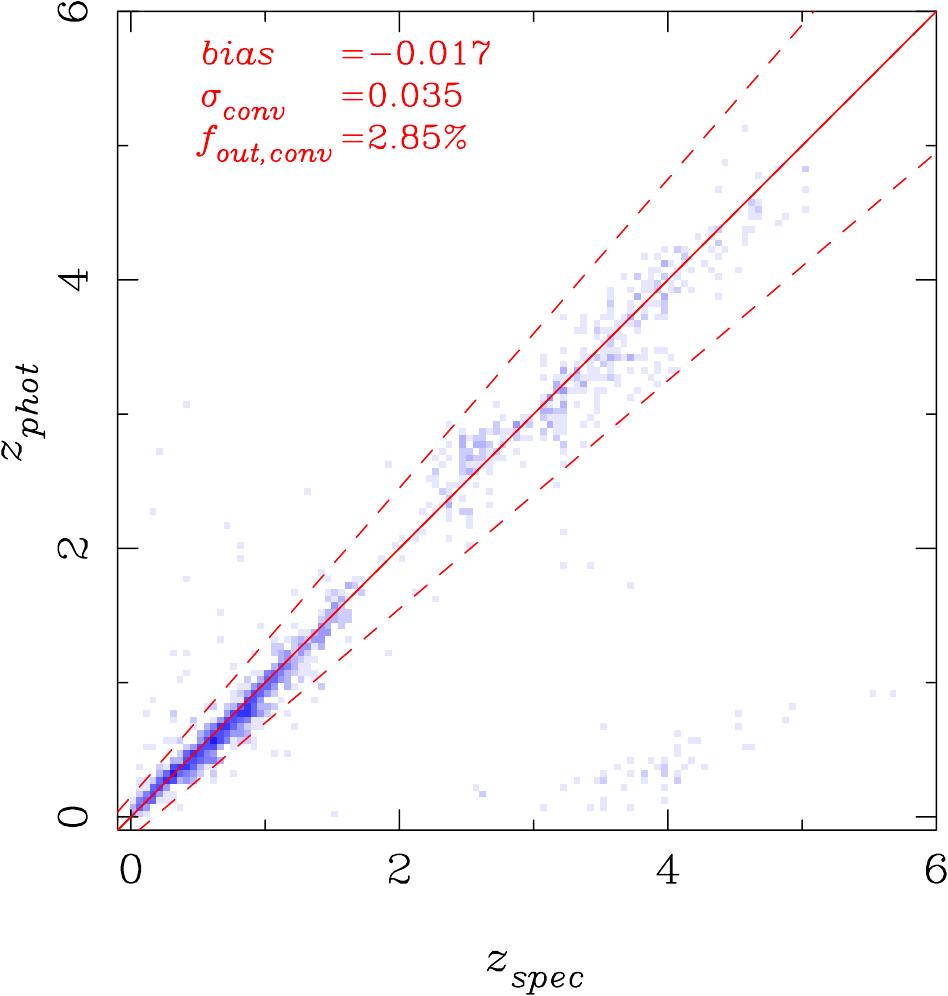}
\caption{
    \textit{Left}: Comparison between our photo-$z$'s and the archival spec-$z$'s in the COSMOS field. A solid line is a one-to-one relation. Objects not between dashed lines are the outliers (i.e., $|z_\mathrm{phot}-z_\mathrm{spec}|/(1+z_\mathrm{spec})>0.15$). Our photo-$z$'s are quite accurate: a small scatter $\sigma_\mathrm{conv}=0.023$, a low outlier rate $f_\mathrm{out,conv}=3.14\%$, and a small systematic offset of $-0.006$.
    \textit{Right}: As in the left panel but for VISTA-VIDEO. The meanings of the symbols are the same. 
}
\label{fig:photoz}
\end{figure*}

Finally, as a quick check of our photo-$z$ accuracy, we compare our photo-$z$'s against spec-$z$'s available in COSMOS, as shown in Fig.~\ref{fig:photoz} (left panel). Spectroscopic objects are a heterogeneous collection of bright objects used to calibrate photo-$z$'s in \citet{weaver20}, and thus, the comparison here is not completely fair, but it is still a useful check.
In brief, our photo-$z$'s are very reliable with a small scatter of $\sigma_\mathrm{conv}=0.023$ and a low outlier\footnote{We define photo-$z$ outlier are those with $|z_\mathrm{phot}-z_\mathrm{spec}|/(1+z_\mathrm{spec})>0.15$} rate of $f_\mathrm{out,conv}=3.14\%$ against spec-$z$'s, where $\sigma_\mathrm{conv}=1.48\times \mathrm{median}[|z_\mathrm{phot}-z_\mathrm{spec}|/(1+z_\mathrm{spec})]$. The systematic offset is very small, at $-0.006$.

Our multi-band catalog is spatially inhomogeneous because it is a collection of data from various surveys. As our focus here is distant clusters, we construct a sub-catalog of homogeneous data in terms of both filter set and spatial coverage while keeping sufficient depth for our goal. We base our analyses on the VISTA-VIDEO region, where we have the deep imaging data from the $u^*$-band to the $K_\mathrm{s}$-band, covering $\sim3.5\, \mathrm{deg}^{2}$. Table~\ref{tab:data} shows a summary of the photometry and the $5\sigma$ limiting magnitude in each band measured with a $1.5\arcsec$ aperture. Although spec-$z$'s available in VISTA-VIDEO are not as extensive as COSMOS, we cross-match our photo-$z$'s with spec-$z$'s from VANDELS \citep{pentericci18,garilli21}, OzDES \citep{lidman20}, and C3R2 \citep{masters17,masters19}. The comparison is shown in the right panel of Fig.~\ref{fig:photoz}. While direct comparisons with COSMOS cannot be made, we observe fairly similar photo-$z$ accuracy despite VISTA-VIDEO being shallower than UltraVISTA.  We are missing spectroscopic objects at $z\sim2$ in VISTA-VIDEO, but given the similarity between COSMOS and VISTA-VIDEO, it is reasonable to assume similarly good accuracy there as well. Again, we focus on the VISTA-VIDEO region in this paper, but a forthcoming paper will present a more extensive cluster search over the wider D/UD fields. 

\begin{figure}
\centering
\includegraphics[width=80mm]{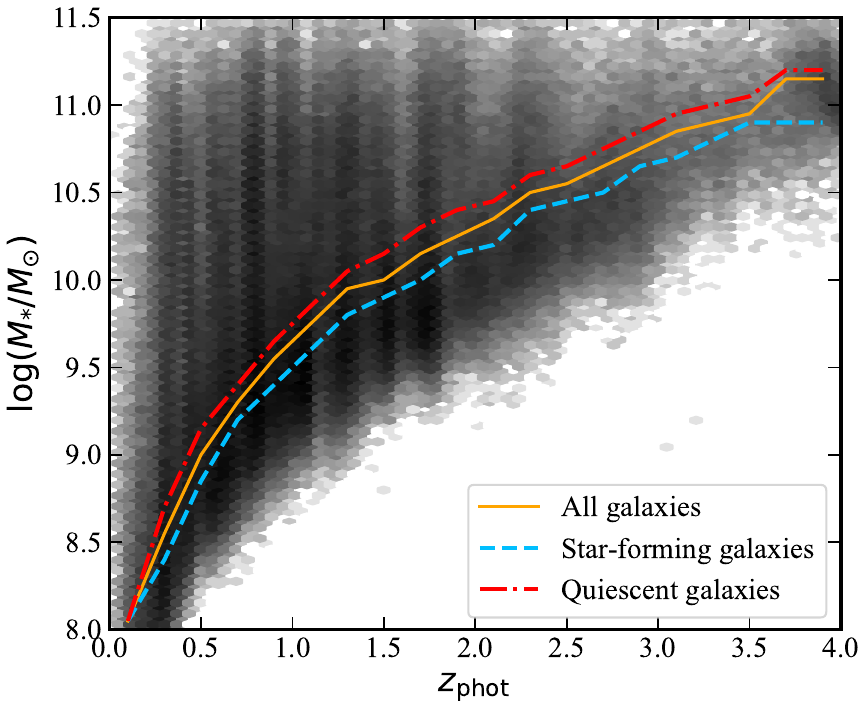}
\caption{
    The stellar mass completeness corresponds to $K_\mathrm{s,lim}=23.6$ estimated by the empirical method introduced in \citet{pozzetti10}. The 90\% complete masses are shown separately for three cases: all galaxies (orange solid line), star-forming galaxies (blue dashed line), and quiescent galaxies (red dotted-dashed line). The background is the two-dimensional histogram for all galaxies. 
}
\label{fig:completeness}
\end{figure}

We then estimate the stellar mass limit of our sample in the VISTA-VIDEO region following an empirical method (\eg, \citealp{pozzetti10,weaver20}). We first calculate the rescaled stellar mass, $M_\mathrm{*,res}$, of each galaxy, 
\begin{equation}
    \log(M_\mathrm{*,res}) =\log(M_{*})-0.4(K_\mathrm{s,lim}-K_\mathrm{s,corr}),
\end{equation}
where $K_\mathrm{s,corr}$ and $K_\mathrm{s,lim}$ are the aperture loss-corrected $K_\mathrm{s}$-band magnitude and the assumed limiting magnitude, respectively. The aperture loss depends on the sizes of objects, but it is typically $0.2$--$0.5$ mag for faint ($K_\mathrm{s}\sim24$) objects. Therefore, we conservatively adopt $K_\mathrm{s,lim}=23.6$. Then, we define the stellar-mass completeness limit as the $90^\mathrm{th}$ percentile of $\log(M_\mathrm{*,res})$ distribution at a given redshift. Fig.~\ref{fig:completeness} shows limiting stellar mass at 90\% completeness calculated above. 
Regardless of the star formation category, galaxies that are more massive than $10^{10.5}\,M_{\odot}$ at $z\sim2$ are almost completely detected.
Then, we also calculate limiting stellar masses for star-forming galaxies and quiescent galaxies separately. Here, we define quiescent galaxies as those whose specific-SFR ($\mathrm{SFR}/M_{*}$; sSFR) are by $\sim1$ dex lower than the star formation main sequence (see eq.~(12) of \citealp{tanaka15}). 
To be specific, we define quiescent galaxies at $z\sim2$ as those whose $\mathrm{sSFR}<10^{-10}\,\mathrm{yr}^{-1}$. 
We find the limiting masses of quiescent and star-forming galaxies at $z\sim2$ to be $\sim 10^{10.5} M_\odot$ and $\sim 10^{10.3}M_\odot$, respectively.

\section{Analysis and results} \label{sec:analysis}
\subsection{Analysis} \label{subsec:analysis}
We use the photo-$z$ catalog constructed in the previous section to search for distant ($z\gtrsim2$) clusters. 
Most (proto)cluster searches in the distant Universe have used star-forming galaxies as a tracer (\eg, \citealt{toshikawa14, toshikawa18}). However, recent observations imply that there are systems with a large fraction of quiescent galaxies \citep{mcconachie22, ito22}, which may be missed when star-forming galaxies are used. We therefore adopt an approach complementary to previous works and use quiescent galaxies as a tracer. 

For our goal, we need a clean and complete quiescent galaxy sample with reliable photo-$z$'s. We therefore adopt the following criteria to select quiescent galaxies: 
\begin{enumerate}
    \item sSFR $ < 10^{-10} \rm ~yr^{-1}$,
    \item $M_\ast > 3 \times 10^{10} ~M_\odot$,
    \item $|z_\mathrm{phot}-z_\mathrm{ref}|<0.2$,
    \item $(z_\mathrm{68max} - z_\mathrm{68min})/(1 + z_\mathrm{phot}) < 0.2$,
    \item $\chi_\nu^2 < 5$,
    \item $K_{\mathrm{s},\mathrm{corr}} < 23.6 ~\mathrm{mag}$,
\end{enumerate}
where $z_\mathrm{ref}$ is the reference redshift we explore, 
$\chi^2_\nu$ is the reduced chi-square of the best-fit model from our photo-$z$ code, and
$z_\mathrm{68min}$ and $z_\mathrm{68max}$ are the lower and upper range of the 68\% confidence interval. In this paper, we focus on $z_\mathrm{ref} = 2.1$, where we identify several good candidates. As noted earlier, we present extensive work over a wider redshift range in a future paper. Note as well that the photo-$z$ accuracy in the left panel of Fig.~\ref{fig:photoz} at $1.9<z_{spec}<2.3$ is $\sigma_{conv}=0.061$ and an outlier fraction of 5.4\%.

After selecting the quiescent galaxies, we define 2D grids with a $10\arcsec$ interval covering the entire XMM-LSS field and estimate the quiescent galaxy number density at each grid. 
We use the kernel density estimate with a 2D Gaussian kernel $K(r)$, which is expressed as
\begin{equation}
    K(r) \propto \exp{\left(-\frac{r^2}{2h^2}\right)},
\end{equation}
where $r$ is the projected distance between a given grid and a galaxy, and $h$ is the Gaussian kernel width parameter. 
We choose the Gaussian kernel widths $h$ of $10\arcsec$ and $60\arcsec$, corresponding to $\sim80$ and $\sim500$ physical kpc at $z=2$, respectively. 
While the $10\arcsec$ kernel is sensitive to the small-scale concentrations of quiescent galaxies, the $60\arcsec$ kernel may well trace more extended structures comparable to halo size. Each of these kernels is subject to contamination by spurious over-densities; the small kernel shows a significant over-density where only a couple of galaxies are located close to each other, and the large kernel is sensitive to a loose concentration of galaxies without an obvious central galaxy. We find that the combination of density maps with these two different scales is efficient in removing such spurious systems. We thus calculate the significance of the kernel density at each grid position against whole density distributions with $h=10\arcsec$ ($\sigma_{10\arcsec}$) and $h=60\arcsec$ ($\sigma_{60\arcsec}$). 
The significance of each grid ($\sigma_h$) is defined as
    \begin{equation}
        \sigma_h = \frac{\rho^h_i - <\rho_h>}{\mathrm{S.D.}},
    \end{equation}
where $\rho^h_i$ is the kernel number density of quiescent galaxies measured at the $i$-th grid with the kernel width of $h$. $<\rho_h>$ and S.D. are the mean and standard deviation of $\rho^h_i$ distribution, respectively. 
Finally, we multiply the values of $10\arcsec$ kernel significance and those of $60\arcsec$ at each grid.

\begin{figure*}
\plotone{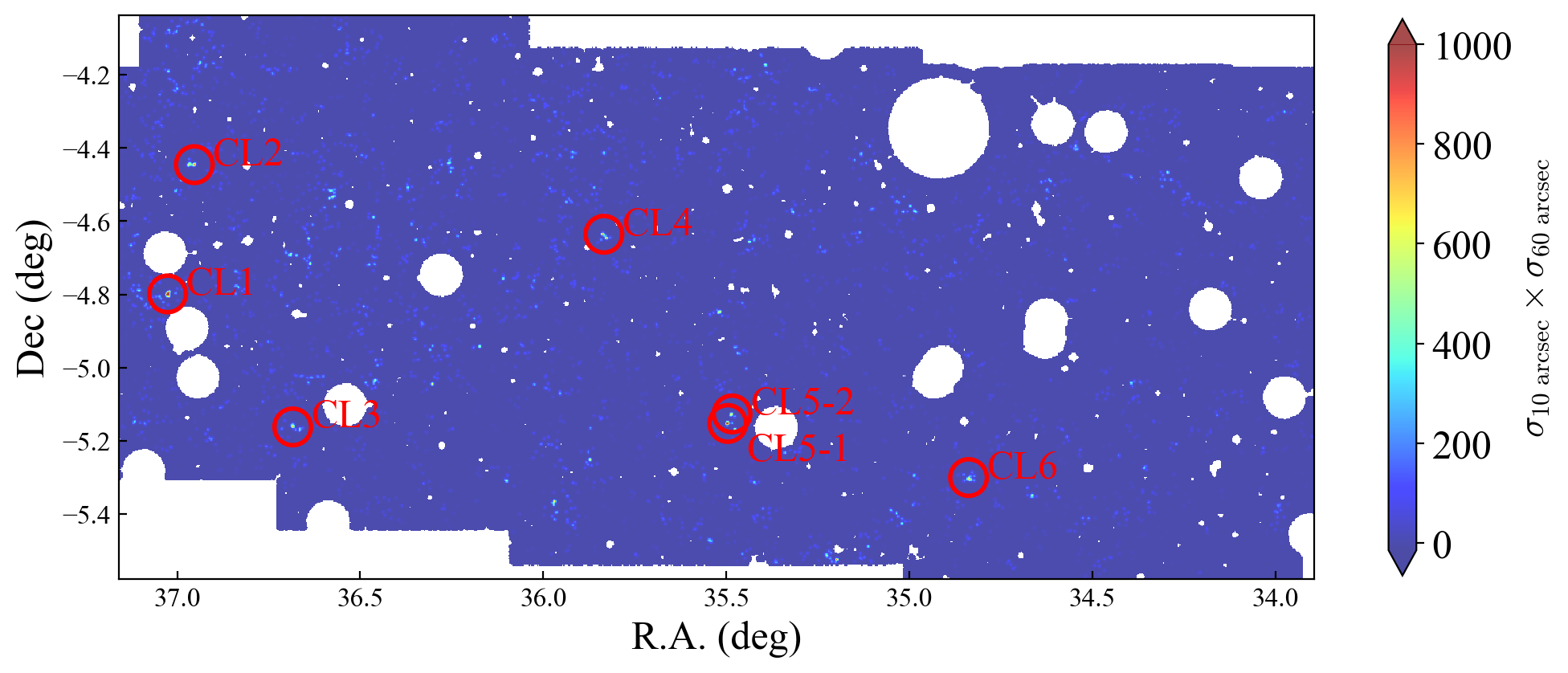}
\caption{Overdensity map ($\sigma_{10\arcsec}\times \sigma_{60\arcsec}$; kernel density estimate with Gaussian kernel and multiplying the results of $10\arcsec$ and $60\arcsec$ kernel width) in the range of $1.9<z_\mathrm{phot}<2.3$ at XMM-LSS field. Higher (lower) significance regions are shown in redder (bluer). The positions of the cluster candidates are shown in red circles. 
\label{fig:densmap}}
\end{figure*}

The color-coded multiple density significance map ($\sigma_{10\arcsec}\times \sigma_{60\arcsec}$) in the XMM-LSS field is shown in Fig.~\ref{fig:densmap}. There are several high-density peaks. Since we aim to construct a clean cluster sample rather than a complete one, we set a fairly strict border value, $\sigma_{10\arcsec}\times \sigma_{60\arcsec}=600$, to regard density peaks as cluster candidates. 11 density peaks remain above this threshold. We then visually inspect the images around these candidates and remove spurious systems: those with false-detected objects around bright galaxies (2 candidates) and extremely concentrated mergers without spatially loose extended structures (2 candidates), which boost $\sigma_{10\arcsec}$. After this screening, we finally select seven cluster candidates at $z\sim2$, shown by the red circles in Fig.~\ref{fig:densmap}. We note that if we adopt the lower $\sigma_{10\arcsec}\times \sigma_{60\arcsec}$ threshold than 600, the number of spurious systems increases. This threshold is a trade-off between completeness and quality. Spectroscopic follow-up observations are needed to validate it. 

In general, different density measures trace slightly different structures \citep{muldrew12}. As a complementary check, we test whether our cluster candidates remain significant if we adopt other density measures. We adopt one of the widely used density measures, 10th-nearest neighbor density, and found that all our candidates are $>30\sigma$. This suggests that our cluster candidates are actually rich and associated with a sufficient number of massive quiescent members. 

As a further check, we also perform our density measurements at $z_\mathrm{ref} = 1.8$ and $1.6$, where known clusters are located in the same XMM-LSS field. At $z_\mathrm{ref} = 1.8$, we successfully identify a massive cluster at $z_\mathrm{spec}=1.80$ (\citealt{newman14}) with $\sigma_{10\arcsec}\times \sigma_{60\arcsec} = 1400$ ($\sigma_{10\arcsec}= 127$ and $\sigma_{60\arcsec} = 11$). Similarly, the well-known cluster at $z_\mathrm{spec}=1.62$ (\citealt{tanaka10, papovich10}) is recovered with $\sigma_{10\arcsec}\times \sigma_{60\arcsec} =706$ ($\sigma_{10\arcsec}= 85$ and $\sigma_{60\arcsec} = 8.3$) 
at $z_\mathrm{ref}=1.6$. These results determine the reliability of our cluster selection procedure. 

\subsection{Cluster candidates at $z\sim2$} \label{subsec:candidates}
\begin{figure*}
    \gridline{\fig{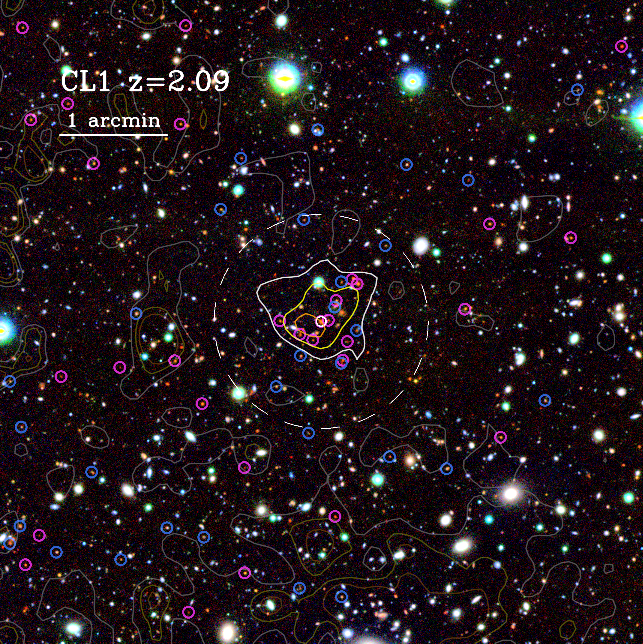}{0.35\textwidth}{}
              \fig{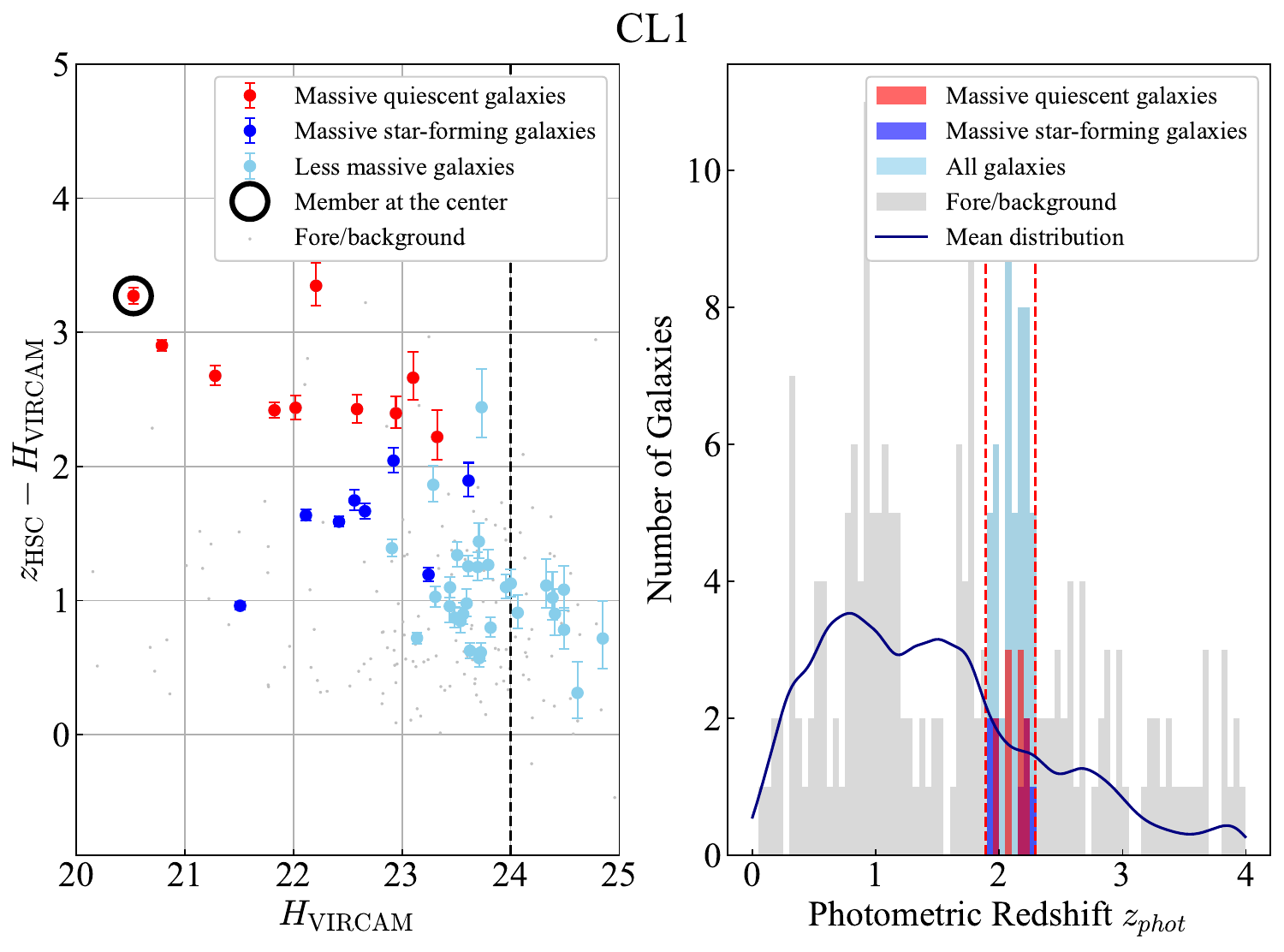}{0.5\textwidth}{}
              }
    \caption{
    \textit{Left}: The pseudo-color image of CL1 (blue: Hyper Suprime-Cam/$z$-band, green: VIRCAM/$J$-band, red: VIRCAM/$K_\mathrm{s}$-band). 
    Galaxies with red and blue circles are massive quiescent ($M_* > 3 \times 10^{10} ~M_\odot, ~\mathrm{sSFR} < 10^{-10} \rm ~yr^{-1}$) and massive star-forming ($M_* > 3 \times 10^{10} ~M_\odot, ~\mathrm{sSFR} > 10^{-10} \rm ~yr^{-1}$) galaxies at $|z_\mathrm{cluster} - z_\mathrm{phot}| < 0.2$, respectively.
    The white circle highlights the most massive galaxy in the cluster candidate. 
    The redshift of this candidate is shown at the top left corner, which is defined as the median redshift of massive quiescent galaxies within $1\arcmin$ (white dashed circle) from the cluster center (i.e. the position of the most massive galaxy). 
    The contours show X-ray emission (0.5--2 keV of XMM-SERVS; \citealt{chen18}) at 1$\sigma$ (white), 3$\sigma$ (yellow), 5$\sigma$ (orange) significance. The contours within $1\arcmin$ from the cluster center are emphasized with thick lines. 
    \textit{Middle}: The color-magnitude diagram (CMD). The red, blue, and light blue symbols represent massive quiescent, massive star-forming, and less massive ($M_* < 3 \times 10^{10} ~M_\odot$) galaxies, respectively, within $1\arcmin$ from cluster center and redshift slice of $|z_\mathrm{cluster} - z_\mathrm{phot}| < 0.2$. The gray dots are those within $1\arcmin$ from the cluster center but out of $|z_\mathrm{cluster} - z_\mathrm{phot}| < 0.2$. The black circle highlights the member galaxy at the cluster center (i.e., the most massive galaxy). The vertical dotted line infers the $H_\mathrm{VIRCAM}$ band magnitude limit.
    \textit{Right}: The redshift distribution of galaxies around the cluster candidate. The meaning of each color is the same for the CMD except for the light blue bins, which show all galaxies within $1\arcmin$ from cluster center and redshift slice of $|z_\mathrm{cluster} - z_\mathrm{phot}| < 0.2$. 
    The solid blue line is the expected number of galaxies within $1\arcmin$, which is estimated by scaling the photo-$z$ distribution of galaxies in the VISTA-VIDEO region to a $1\arcmin$ aperture. Note the significant excess of galaxies around the cluster redshift. 
    }
\label{fig:image1}
\end{figure*}

\begin{figure*} 
    \gridline{\fig{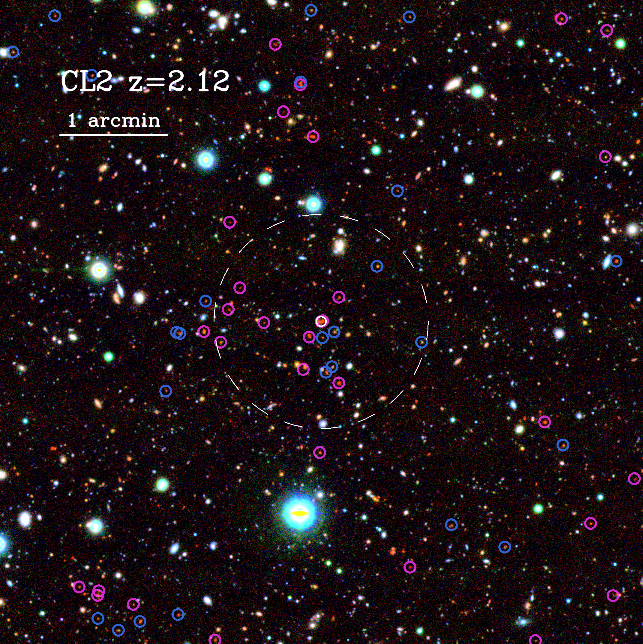}{0.35\textwidth}{} 
              \fig{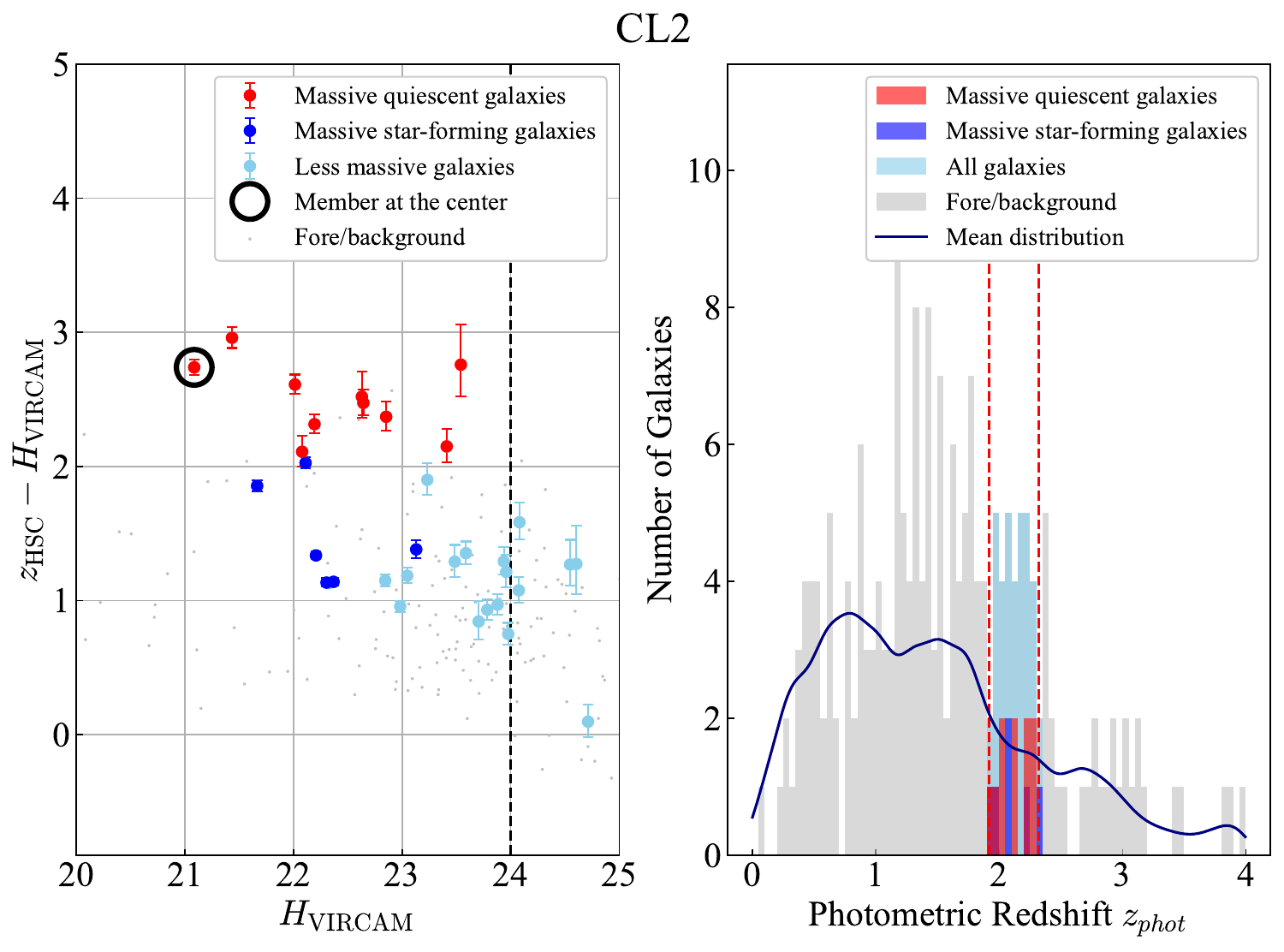}{0.5\textwidth}{}
              }
    \gridline{\fig{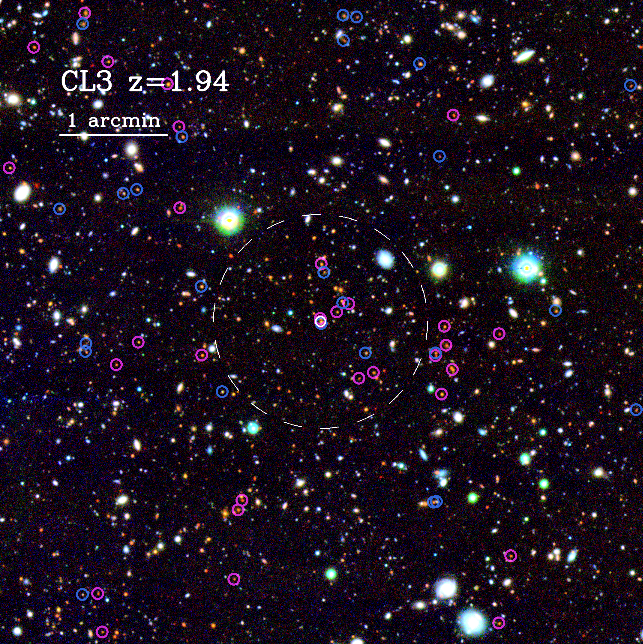}{0.35\textwidth}{} 
              \fig{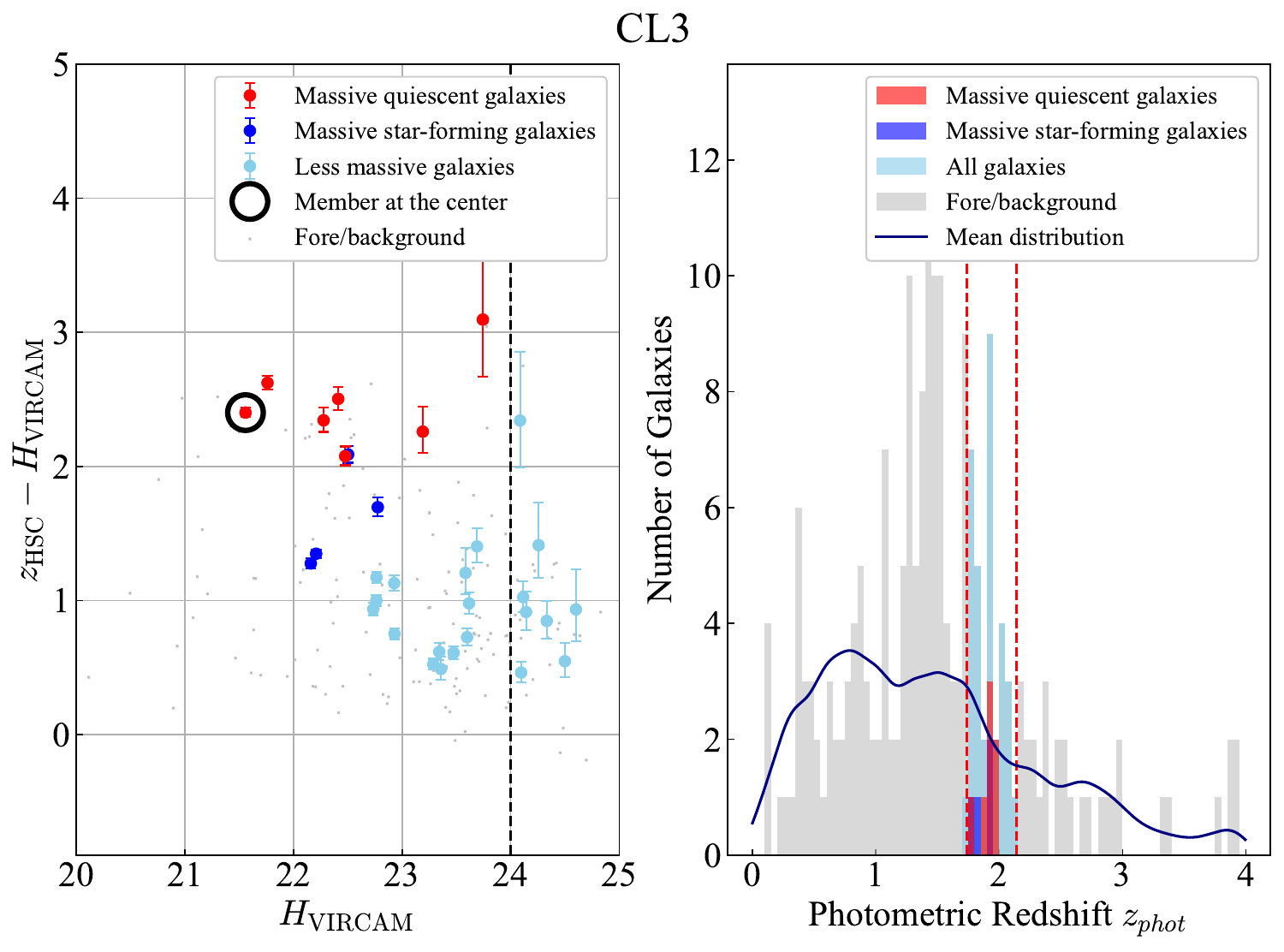}{0.5\textwidth}{}
              }
    \gridline{\fig{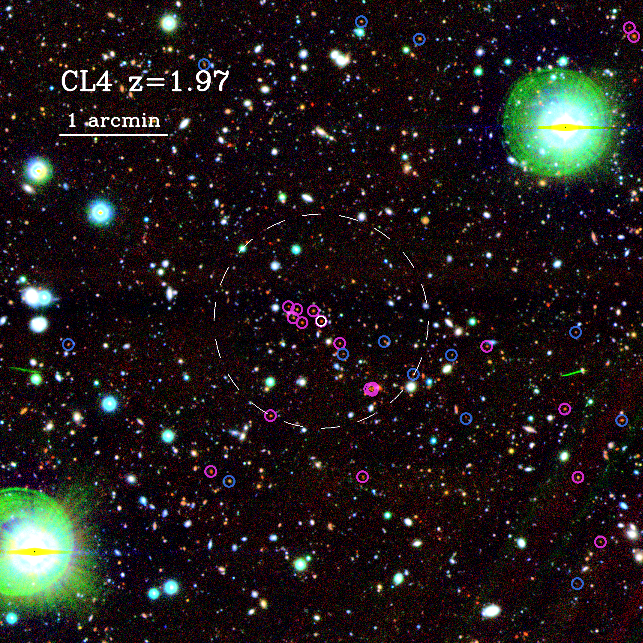}{0.35\textwidth}{}
              \fig{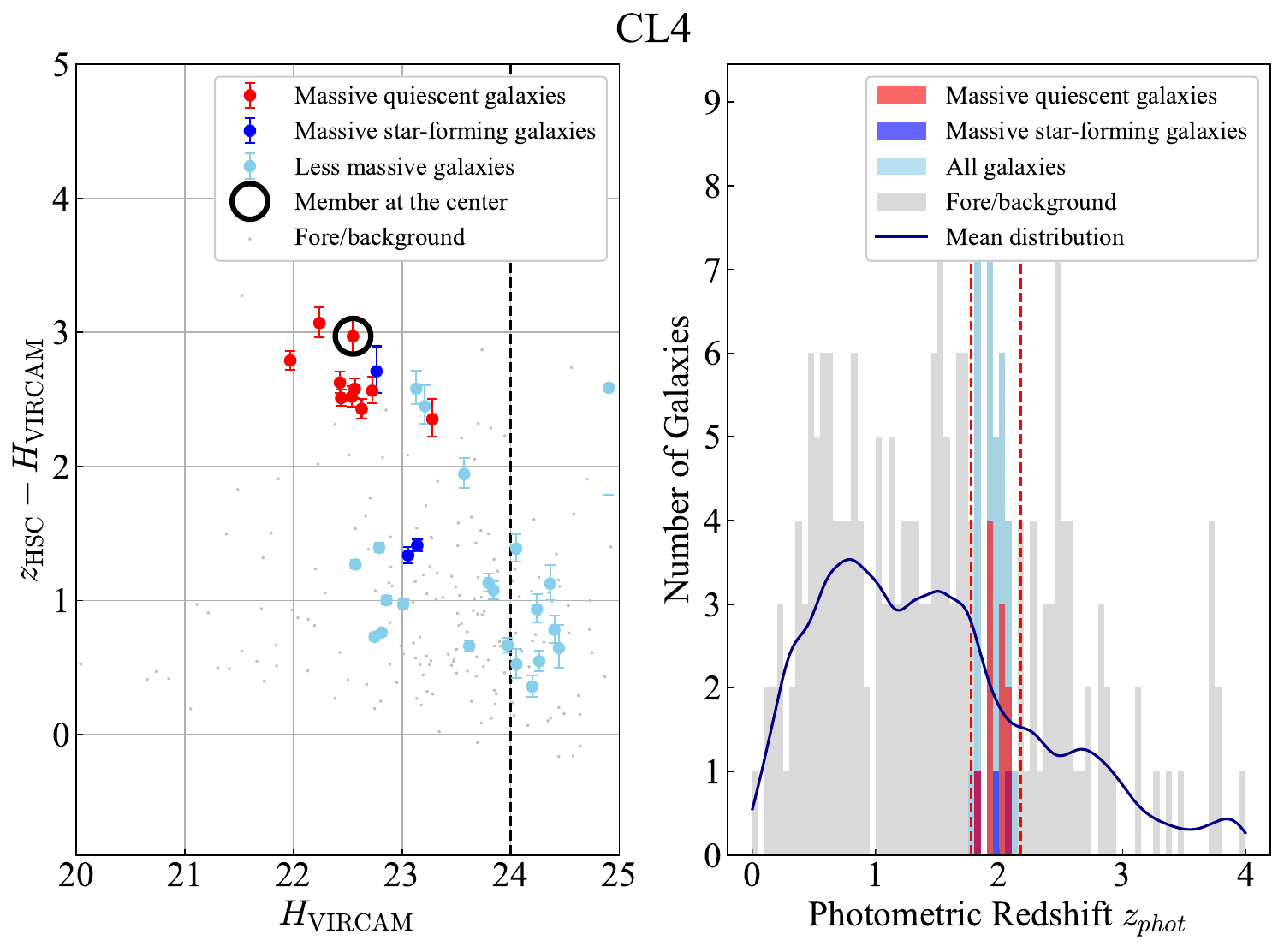}{0.5\textwidth}{}
              }
    \caption{The same as in Fig.~\ref{fig:image1} but for CL2, CL3, and CL4. }
\label{fig:image2}
\end{figure*}

\begin{figure*}
    \gridline{\fig{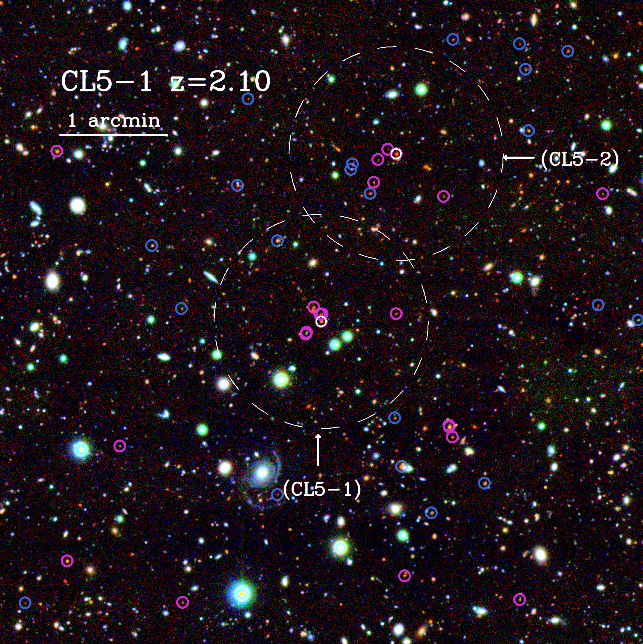}{0.35\textwidth}{}
              \fig{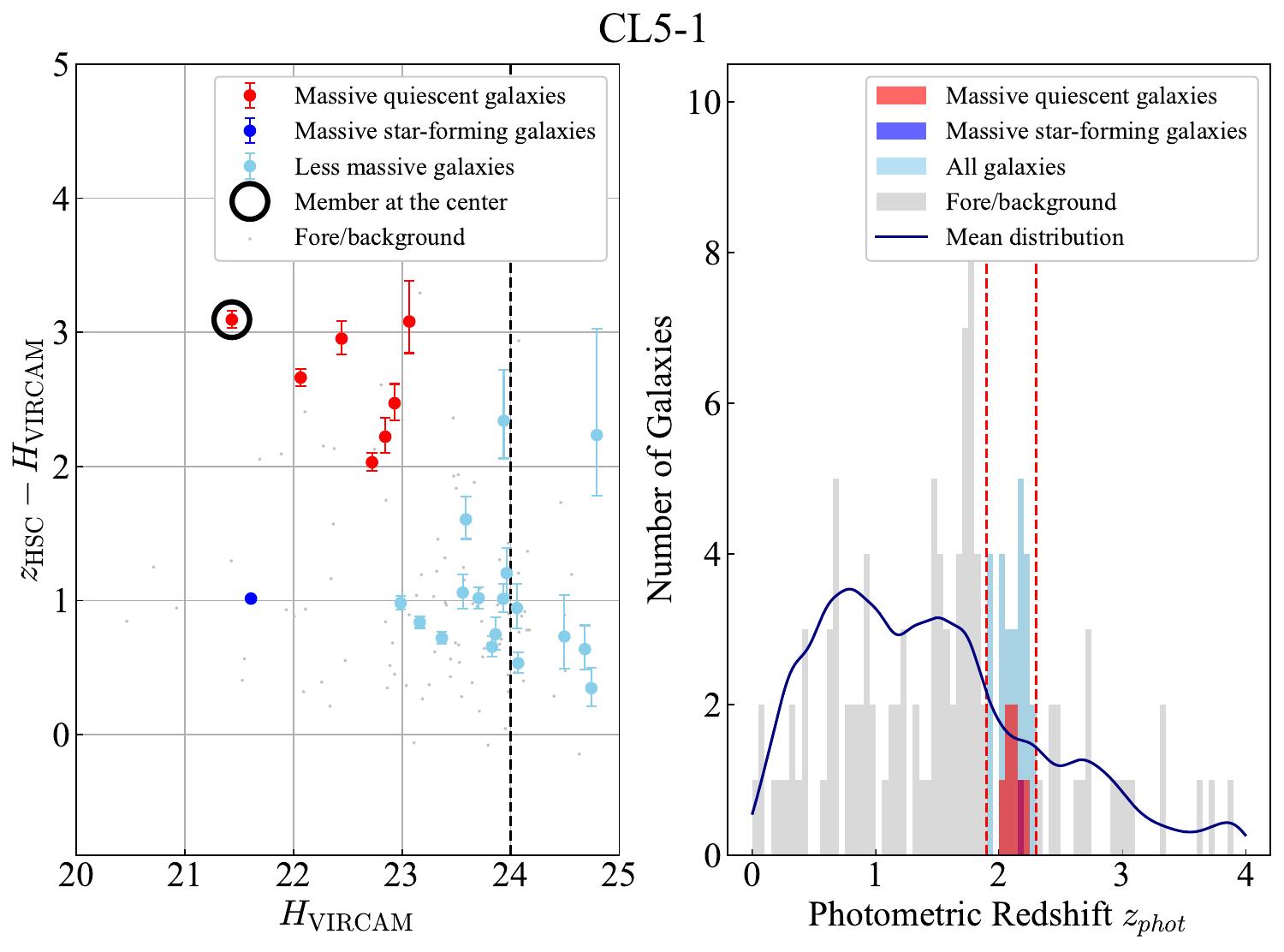}{0.5\textwidth}{}
              }
    \gridline{\fig{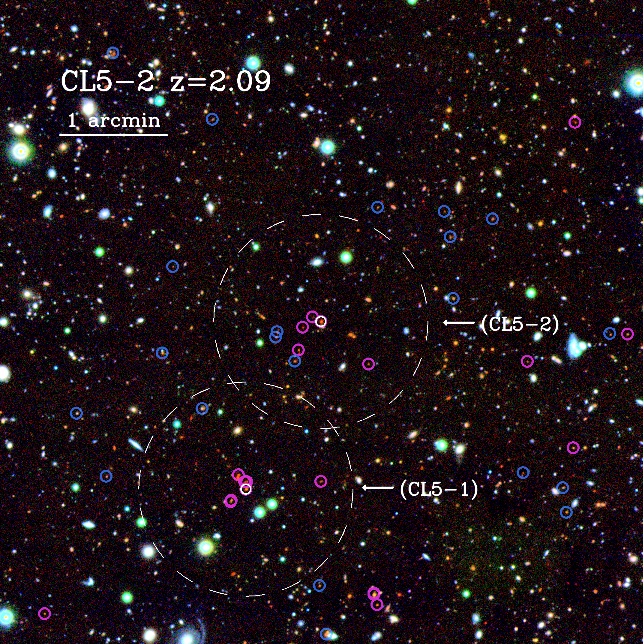}{0.35\textwidth}{}
              \fig{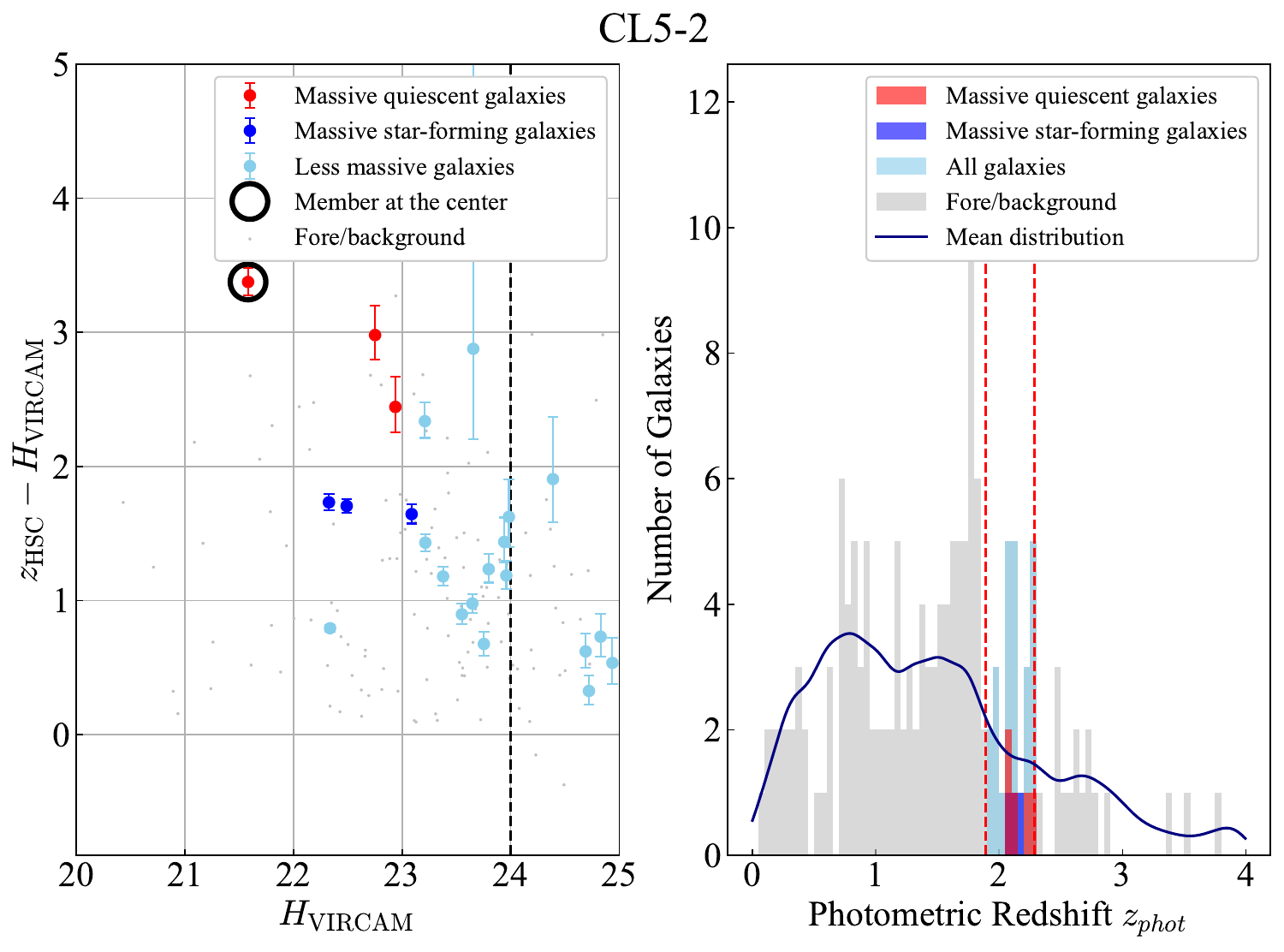}{0.5\textwidth}{}
              }
    \gridline{\fig{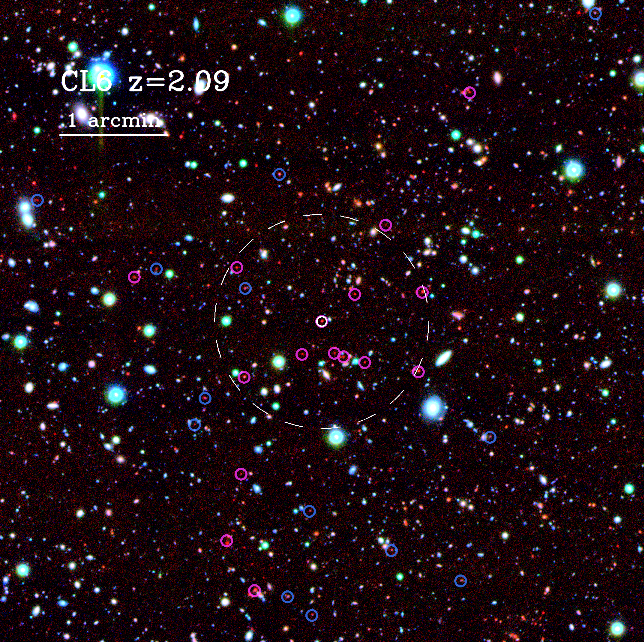}{0.35\textwidth}{}
              \fig{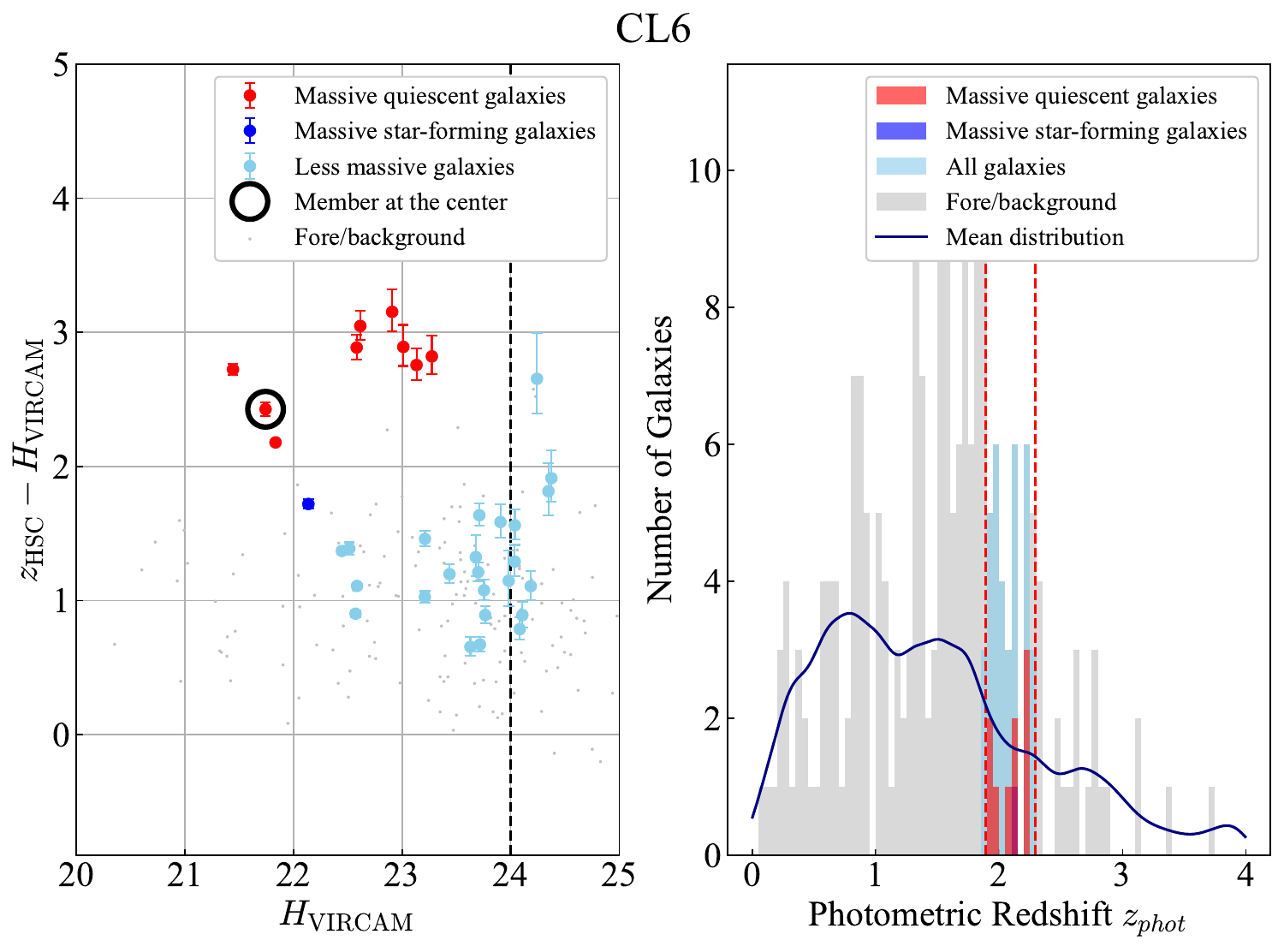}{0.5\textwidth}{}
              }
    \caption{The same as in Fig.~\ref{fig:image1} but for CL5-1, CL5-2, and CL6.}
\label{fig:image3}
\end{figure*}

We identify seven prominent overdensity peaks of massive quiescent galaxies at $z \sim 2$ as cluster candidates in the XMM-LSS field. All candidates satisfy $\sigma_{10\arcsec}\times \sigma_{60\arcsec} > 600$ and in detail, $\sigma_{10\arcsec} \gtrsim 6$ and $\sigma_{60\arcsec} \gtrsim 100$. We define the center of each cluster candidate as the position of the most massive galaxy around the overdensity. We also define a redshift of a given candidate ($z_\mathrm{cluster}$) as the median of photo-$z$ values of the massive quiescent galaxies within $1\arcmin$ (corresponding to $0.5$ physical Mpc at $z \sim 2$) from the cluster center and satisfy $|z_\mathrm{ref} - z_\mathrm{phot}| < 0.2$ with $z_\mathrm{ref} = 2.1$. Table~\ref{tab:candidates} summarizes the positions and physical properties of the candidates. Figs.~\ref{fig:image1}--\ref{fig:image3} display pseudo-color images of the candidates (left), color-magnitude diagrams (CMDs; $z_\mathrm{HSC}-H_\mathrm{VIRCAM}$ vs. $H_\mathrm{VIRCAM}$; middle), and redshift distributions of galaxies around the candidates (right). 

In the left and right panels of Figs.~\ref{fig:image1}--\ref{fig:image3}, we confirm the spatial and redshift concentrations of massive galaxies (highlighted by red or blue symbols) around the cluster candidates. In addition, we also check the concentration including less massive galaxies. In the right panels, blue solid lines indicate the numbers of total galaxies expected to be found within $1\arcmin$ from the cluster centers. We calculate the redshift distribution of galaxies in the entire XMM-LSS field and normalize it to match the area of a circle with $1\arcmin$ radius. We observe a larger number of galaxies around the cluster redshifts (light blue) than expected, suggesting that massive quiescent galaxies and other galaxies such as low-mass star-forming galaxies are concentrated both in space and redshift. 

The CMDs of these candidates show clear red sequences (middle panels of Figs.~\ref{fig:image1}--\ref{fig:image3}). This indicates that evolved quiescent galaxies populate in the cluster candidates. Furthermore, the number of massive quiescent galaxies is larger than massive star-forming galaxies. 
This implies that massive galaxies in our cluster candidates tend to be quiescent rather than actively star-forming, which is opposite to bursty star-forming (proto)clusters reported at $z\gtrsim2$ (\eg, \citealp{wang16,miller18,oteo18}). We further explore this trend in the context of environmental quenching in Section~\ref{sec:discussion}. 

Interestingly, CL5-1 and CL5-2 are close to each other on the sky: the projected distance between them is $\sim 1\arcmin.7$ (Fig.~\ref{fig:image3}). Since they have almost the same redshift ($z_\mathrm{phot}=2.10$ and $z_\mathrm{phot}=2.09$), CL5-1 and CL5-2 might be pair-clusters before merging. 

We note that CL6 is located in the region, where the $J$-band is missing. CL6 is detected by the galaxies satisfying all criteria described above (Section~\ref{subsec:analysis}), and the physical properties of the member galaxies might be well determined. However, at $z\sim2$, the $J$-band ($\sim 1.2~\micron$) is important to capture the Balmer/4000~\AA~ break and constrain the spectral energy distribution. Therefore, to be conservative, we exclude CL6 from the cluster sample in the discussion section (Section \ref{sec:discussion}). 

    \begin{deluxetable*}{lccccccl}
        \tablecaption{Physical properties of cluster candidates \label{tab:candidates}}
        \tablewidth{0pt}
        \tablehead{
        \colhead{ID} & \colhead{R.A.} & \colhead{Dec.} & \colhead{$z_\mathrm{phot}$} & \colhead{$L_\mathrm{X}$} & \colhead{$M_{200}$} & $\sigma_\mathrm{X}$ & \colhead{Comments}\\
        \colhead{} & \colhead{(h m s)} & \colhead{(d m s)} & \colhead{} & \colhead{($\rm 10^{44}~erg~s^{-1}$)} & \colhead{($10^{13}~M_\odot$)} & \colhead{} & \colhead{}
        }
        \decimalcolnumbers
        \startdata
        CL1 & 02$^h$ 28$^m$ 06.29$^s$ & $-04$\textdegree $47\arcmin55\arcsec.6$ & 2.09 & $1.46 \pm 0.35$ & $7.75 \pm 1.15$ & 4.14 & extended X-ray emission \\
        CL2 & 02$^h$ 27$^m$ 48.80$^s$ & $-04$\textdegree $26\arcmin44\arcsec.1$ & 2.12 & $0.84 \pm 0.40$ & $5.34 \pm 1.50$ & 2.12 \\
        CL3 & 02$^h$ 26$^m$ 44.27$^s$ & $-05$\textdegree $09\arcmin43\arcsec.7$ & 1.94 & $<$0.45 & $<$4.09 & 0.83 \\
        CL4 & 02$^h$ 23$^m$ 20.31$^s$ & $-04$\textdegree $38\arcmin08\arcsec.9$ & 1.97 & $<$0.82 & $<$6.02 & 1.74 \\
        CL5-1 & 02$^h$ 21$^m$ 59.10$^s$ & $-05$\textdegree $09\arcmin09\arcsec.6$ & 2.10 & $<$1.16 & $<$6.87 & 1.52 & separation of CL5-1 and CL5-2 are $\sim1\arcmin.7$\\
        CL5-2 & 02$^h$ 21$^m$ 56.28$^s$ & $-05$\textdegree $07\arcmin35\arcsec.7$ & 2.09 & $<$1.22 & $<$7.10 & 1.89 \\
        \hline
        CL6$^\dagger$ & 02$^h$ 19$^m$ 21.26$^s$ & $-05$\textdegree $18\arcmin01\arcsec.7$ & 2.09 & $<$0.51 & $<$3.98 & 0.84 & $J$ band missing candidate \\
        \enddata
        \tablecomments{
        (1) cluster candidate ID; 
        (2) \& (3) coordinates (the positions of the massive member as indicated by the white circles in Fig.~\ref{fig:image1}--\ref{fig:image3}); 
        (4) photometric redshift (median photo-$z'$ of the massive quiescent galaxies within $1\arcmin$ from the most massive member as shown by the white dotted circle in Fig.~\ref{fig:image1}--\ref{fig:image3}); 
        (5) X-ray luminosity in $0.1$--$2.4$ keV band (point sources are removed.);  
        (6) virial mass of $M_{200}$; 
        (7) X-ray flux significance; 
        (8) the properties of each candidate. \\
        CL3--CL6 have under 2$\sigma$ significance in X-ray flux, so we show the 2$\sigma$ upper limits of their X-ray luminosity and $M_{200}$. \\
        $^\dagger$: CL6 is in $J$-band missing area, so we exclude it from our sample in the discussion (Section \ref{sec:discussion}) conservatively (see Section~\ref{subsec:candidates}).}
    \end{deluxetable*}


\subsection{X-ray data from XMM-Newton}
\label{subsec:xray}
An extended X-ray emission is an excellent tracer of massive gravitationally-collapsed systems. We use the archival data of XMM-Newton to search for extended X-ray emission. Following the recipe of \citet{finoguenov10}, we measure X-ray flux using 24$\arcsec$ radius apertures in the 0.5--2.0 keV after removing the instrumental backgrounds, unresolved sky backgrounds, and all the detected point sources. To convert from the observed 0.5--2.0 keV band flux to the rest-frame 0.1--2.4 keV band flux, we perform the $k$-correction based on the estimated temperature and redshift. Finally, we estimate the virial mass of $M_{200}$ from the scaling relation presented in \citet{leauthaud10}, which is validated to $z\sim2$ by \citet{bethermin14}. Our results are summarized in Table~\ref{tab:candidates}. 

We clearly detect a significant X-ray emission ($>4\sigma$) from CL1. The emission is extended to $\sim 40\arcsec$ in diameter, corresponding to $\sim350$ physical kpc at $z=2$. In the left panel of Fig.~\ref{fig:image1}, we overlay the contours of the detected X-ray (0.5--2~keV data from XMM-SERVS; \citealt{chen18}) on the pseudo-color image of CL1, which demonstrates the excellent association of the X-ray emission with the galaxy overdensity. The X-ray luminosity is estimated to be $L_\mathrm{X} = (1.46 \pm 0.35) \times 10^{44} \mathrm{~erg ~s^{-1}}$ in the 0.1--2.4~keV band, and the virial mass \footnote{We define $r_{200}$ as the radius within which the mean interior density is 200 times the critical density of the Universe at the cluster redshift. $M_{200}$ is the mass within $r_{200}$.} of $M_{200} = (7.75 \pm 1.15) \times 10^{13} ~M_\odot$. The corresponding virial radius is $r_{200} = 414 ~\mathrm{kpc}$, consistent with the spatial extent of the galaxy overdensity. These results support the idea that CL1 is a well-developed virialized cluster at $z\sim2$. 

Despite the low significance, CL2 is tentatively detected in the X-ray ($\sim2\sigma$) after removing the contribution of point sources (see Section 3.1 of \citealt{finoguenov09} for detailed procedures). Moreover, a point source is detected near the center of CL2. \citet{chen18} have reported the point source (ID XMM05036 of XMM-SERVS point source catalog) located within $2\arcsec$ from the CL2 center. Suppose this point source is associated with the most massive quiescent galaxy of CL2 (white circle in Fig.~\ref{fig:image2}). In that case, it gives us a hint to understand the role of AGN activity for the brightest cluster galaxy (BCG) formation (\eg, \citealt{shimakawa24}). 
However, we cannot conclude which galaxy this point source associates with because of the limited spatial resolution and sensitivity. 

The other five candidates are not detected in X-rays. We just report $2\sigma$ upper limits of their luminosity and virial mass in Table~\ref{tab:candidates}. They might be lower-mass systems below the detection limit. 

We emphasize the importance of a follow-up observation with a high-spatial-resolution X-ray telescope like the Chandra. Although we detect a significant extended emission around CL1, contamination from point sources is still possible. For instance, \citet{logan18} have conducted Chandra follow-up observations for clusters at $z>1$ in the XMM-LSS field and found that some clusters that were previously detected in X-rays by XMM-Newton suffer from contamination by emission from AGNs (see also \citealt{duffy22}). Chandra's high spatial resolution helps us eliminate the AGN contamination and confirm the extended emission around CL1. Moreover, we may be able to locate the detected point source in CL2 to its host by resolving observation.

\subsection{Comparison with previous surveys}
As shown in previous sections, our cluster candidates are plausible in terms of the significant overdensities, the clear red sequence, and the X-ray emission. In the literature, many (proto)cluster surveys have been conducted in the XMM-LSS field, and some have found galaxy overdensities at $z\gtrsim1.5$ (\citealt{willis13, trudeau20, krefting20, gully24}). To see which cluster candidates have been previously reported or are newly identified in this work, we cross-match our candidates with those in the literature. Here, we adopt a matching radius of $2\arcmin$, which is somewhat large compared to cluster size (i.e.,  $1\arcmin$), since some studies focus on such relatively large spatial scales when measuring overdensity. We summarize the cross-match results in Table~\ref{tab:cross-match}. In brief, four out of seven candidates (i.e. CL1, CL3, CL4, and CL6) have at least one counterpart. 

CL1 has been listed in previous studies as a (proto)cluster candidate \citep{trudeau20, gully24}. 
\citet{trudeau20} have reported an extended X-ray source associated with an overdensity of photo-$z$ selected galaxies near CL1. They have found the peak of the photo-$z$ distribution around $z_\mathrm{phot} = 1.79\pm0.14$, slightly lower than our estimate ($z=2.09$). We note, however, that their photo-$z$ distribution has a tail towards higher redshifts up to $z\sim2.5$. \citet{gully24} have found an overdensity of Spizer/Infrared Array Camera (IRAC; \citealt{fazio04}) color-selected sources at $z>1.3$ ($[3.6] - [4.5] > -0.05$) near CL1, although accurate photo-$z$'s are not available. These studies, as well as our rediscovery, strongly support that CL1 is a promising cluster candidate. 

CL3, CL4, and CL6 are matched with the overdensities of \citet{gully24} at separations of $1\arcmin.56$, $1\arcmin.78$, and $1\arcmin.23$, respectively. Considering relatively large spatial separation and redshift uncertainties due to the color selection, it is still unclear whether these overdensities are identical to our cluster candidates at $z\sim2$. CL6 also has a counterpart in \citet{krefting20}, who have searched for overdense regions at $0.1 < z_\mathrm{phot} < 1.67$ using photo-$z$ galaxies with $u$-band to Spitzer/IRAC $4.5\micron$ band photometry. They have found an overdensity in the redshift slice of $1.281 < z <1.665$ located at $1\arcmin.61$ from CL6, which also has a relatively large separation. 

CL2, CL5-1, and CL5-2 have no counterparts in the literature, representing newly identified cluster candidates at $z\sim2$. Moreover, CL3, CL4, and CL6 may also be new clusters due to the ambiguous match with \citet{gully24}. This work demonstrates the possible high efficiency of red-sequence cluster surveys even at high redshift, such as $z\sim2$ (see also \citealt{ito23a, tanaka23}). However, spectroscopic confirmations are urgent for these targets to make a firm conclusion. 

\begin{deluxetable*}{lccccl}
    \tablecaption{Cross-matched candidates with other cluster/group candidates from the literature \label{tab:cross-match}}
    \tablewidth{0pt}
    \tablehead{
    \colhead{ID} & \colhead{$\mathrm{ID}_\mathrm{ref}$} & \colhead{$\Delta r_\mathrm{proj}$} & \colhead{$z_\mathrm{phot,~literature}$} & \colhead{$ z_\mathrm{phot,~this~work}$} & \colhead{references}
    }
    \decimalcolnumbers
    \startdata
    CL1 & 33 & $0\arcmin.13$ & 1.79 & 2.09 & \citet{trudeau20} \\
        & X62 & $0\arcmin.33$ & $>1.3$ & & \citet{gully24}$^\dagger$ \\
    CL3 & X53 & $1\arcmin.56$ & $>1.3$ & 1.94 & \citet{gully24}$^\dagger$ \\
    CL4 & X34 & $1\arcmin.78$ & $>1.3$ & 1.97 & \citet{gully24}$^\dagger$ \\
    CL6 & X18 & $1\arcmin.23$ & $>1.3$ & 2.09 & \citet{gully24}$^\dagger$ \\
        & 322 & $1\arcmin.61$ & $1.281 < z_\mathrm{phot} <1.665$ & & \citet{krefting20} \\
    \enddata
    \tablecomments{
    Summary table of the cross-matched results within $2\arcmin$. 
    (1) our cluster candidate ID; 
    (2) candidate ID presented in references; 
    (3) projected separation between our candidates and those in references; 
    (4) photometric redshift presented in references; 
    (5) photometric redshift estimated in this work;
    (6) reference. \\
    $^\dagger$: \citet{gully24} have estimated candidates' redshifts based on the IRAC color criteria. They have not reported precise photometric redshifts of these candidates.
    }
\end{deluxetable*}

\section{Discussion} \label{sec:discussion}
We have used the deep multi-band imaging data in the XMM-LSS field to search for galaxy clusters at $z\sim2$. Using the high-accuracy photometric redshifts, we have identified seven promising cluster candidates with spatially concentrated massive quiescent galaxies and clear red sequences. One is detected in X-rays, adding further confidence that it is a massive virialized system. Although spectroscopic confirmation is needed to determine cluster membership robustly, we further examine these candidates to investigate the role of the cluster environment on galaxy evolution at this high redshift with the available photometric data. 

In the following subsections, we further discuss the properties of cluster galaxies based on the photo-$z$ sample. To avoid possible contamination of fore/background sources, we subtract the contribution of contaminating galaxies from the galaxy number around each cluster candidate in a statistical manner as: 
\begin{equation}
    N_\mathrm{gal}(z, M_*) = N_\mathrm{C}(z, M_*) - N_\mathrm{survey}(z, M_*)\cdot \frac{A_\mathrm{C}}{A_\mathrm{survey}},
\end{equation}
where $N_\mathrm{gal}(z, M_*)$, $N_\mathrm{C}(z, M_*)$, and $N_\mathrm{survey}(z, M_*)$ are the number of galaxies after correction, the raw count around the cluster candidates within a given aperture with an effective area of $A_\mathrm{C}$, and that in the entire survey field whose effective area is $A_\mathrm{survey}$, respectively, in given redshift and stellar mass bins. 
We set the redshift bin to $|z_\mathrm{phot}-z_\mathrm{ref}|<0.2$ with $z_\mathrm{ref}=2.1$ that is the same value used for the cluster search shown in Section \ref{subsec:analysis}. 
We consider an aperture with a radius of $1\arcmin$ around each cluster candidate (cf. dashed white circles in Figs.~\ref{fig:image1}--\ref{fig:image3}). 

Before we discuss the properties of the cluster galaxies, it is important to note the possible selection bias in our cluster sample. Since we use massive quiescent galaxies as tracers of clusters, our candidates may be biased to be massive mature systems rather than young (unvirialized) systems dominated by star-forming galaxies (\eg, \citealt{toshikawa18}). We aim here to discuss and characterize the environmental effects that quench galaxies at $z\sim2$, complementary to recent studies targeting overdensities of star-forming populations and focusing on the environmentally promoted star formation (\eg, \citealp{wang16, miller18, oteo18, toshikawa24}). For this purpose, our cluster candidates are appropriate targets because environmental quenching effects might be strongest in mature clusters. 

\subsection{Star formation rate and stellar mass relation} \label{subsec:sfr} 
\begin{figure}
\plotone{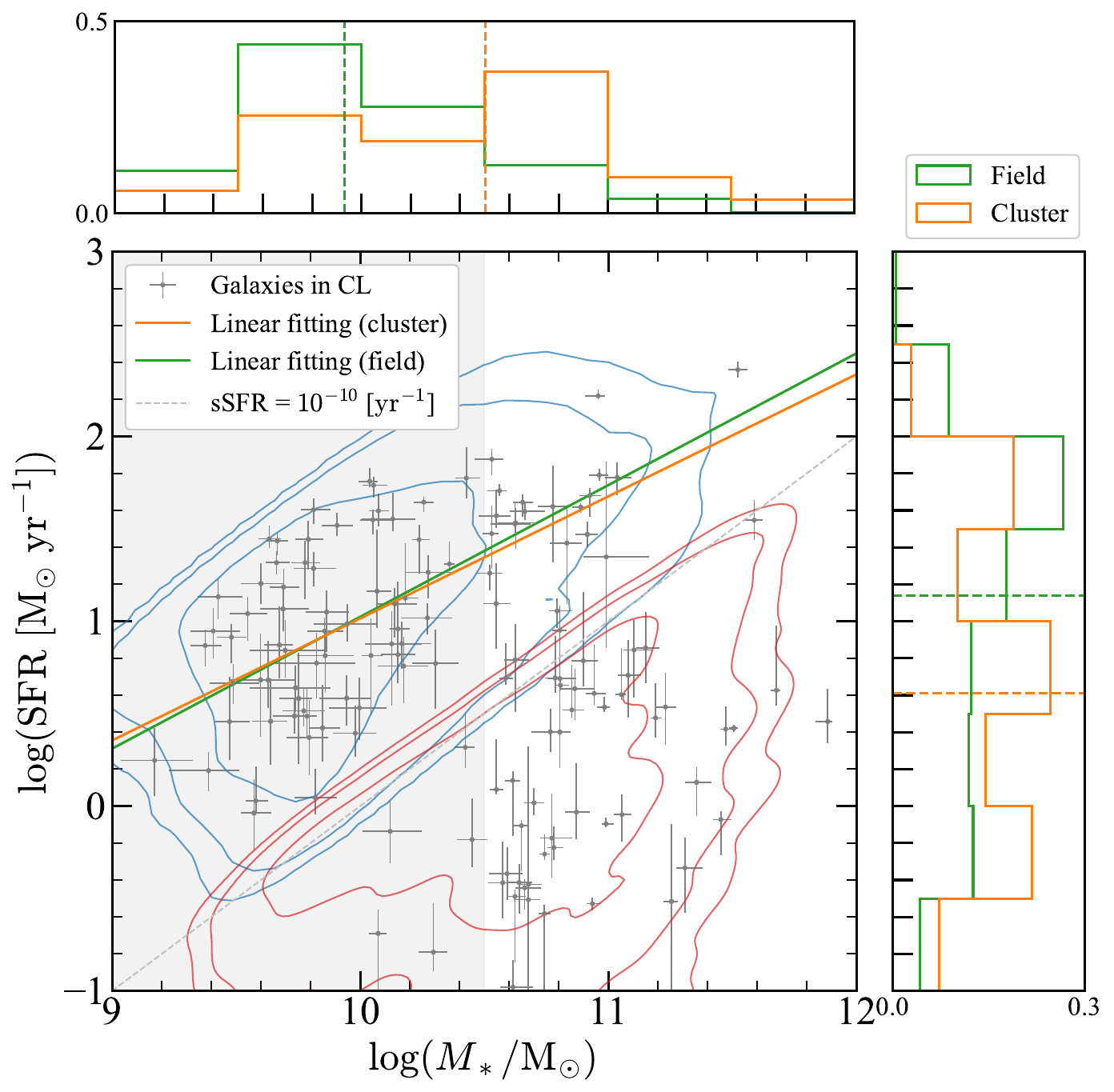}
\caption{
Relation between SFR and stellar mass. 
The blue (red) contours show the distribution of star-forming (quiescent) galaxies in the field at $|z_\mathrm{phot} - 2.1| < 0.2$ with 68\%, 95\%, and 99\%. 
The dotted gray line shows the quiescent criteria we adopt ($\mathrm{sSFR} = 10^{-10} ~\mathrm{yr^{-1}}$). 
The grey points show the member galaxies of the candidates at $|z_\mathrm{phot} - z_\mathrm{cluster}| < 0.2$ after field subtraction and random sampling as a reference. 
The orange (green) line shows the linear least square best-fit result of star-forming galaxies in the cluster candidates (in the field at $|z_\mathrm{phot} - 2.1| < 0.2$). 
The grey shade shows an incomplete mass range. 
The orange and green histograms display the normalized distributions of galaxies in all cluster candidates and field galaxies as a function of stellar mass (top panel) or SFR (right panel). 
In each histogram, the orange and green dotted lines show the median value of the distributions of cluster member galaxies and field galaxies, respectively. The histogram of SFR is only for the galaxies above the stellar mass completeness limit. 
\label{fig:SFR_M}}
\end{figure}

First, we focus on the SFR and stellar mass distributions. Fig.~\ref{fig:SFR_M} shows the cluster galaxy distribution on the SFR versus stellar mass plots (gray dots). Here, for visualization, we show randomly sampled galaxies around clusters to match the expected number of galaxies after the contamination correction (see Section~\ref{sec:discussion}). We overlay blue and red contours to show the distributions of field star-forming and quiescent galaxies, respectively. Star-forming galaxies in the clusters and the field seem to have similar distributions. On the other hand, the cluster candidates are likely to lack low-mass ($\log(M_{*}/M_{\odot})<10.5$) quiescent galaxies compared to the field although this mass range is below the mass limit. Comparing the stellar mass distributions of the clusters and the field for the whole galaxy population (top panel), we find the cluster galaxy distribution to be skewed toward the high-mass regime. This top-heavy stellar mass distribution has been reported in the literature for clusters/groups (\eg, \citealt{vanderburg13, vanderburg18, vanderburg20, ando20, ando2022}), suggesting accelerated stellar mass growth at their formation epoch (\eg, $z>3$). 

It has been reported that the typical SFRs of the star-forming galaxies in clusters at $1<z<1.5$ are lower than those in field \citep{old20}, while \citet{nantais20} have argued no SFR suppression in SpARCS clusters at $z\sim1.6$. To test whether star formation suppression occurs in our cluster candidates, we also compare the positions of the star-formation main sequence in the clusters and the field. We fit $\log(\mathrm{SFR})$--$\log(M_{*})$ relation of star-forming galaxies with a linear function and plot the best-fit lines for the cluster candidates (orange) and the field (green) in Fig.~\ref{fig:SFR_M}. For the cluster galaxies, we repeat the statistical field subtraction mentioned above 100 times and derive the average relation of the 100 best fits to minimize the statistical error in random sampling. The derived main sequences are similar between the clusters ($\log{(\mathrm{SFR}~[M_\odot~\mathrm{yr^{-1}]})} = 0.66\log{(M_*/M_\odot)} - 5.6$) and the field ($\log{(\mathrm{SFR}~[M_\odot~\mathrm{yr^{-1}]})} = 0.71\log{(M_*/M_\odot)} - 6.1$), suggesting that no SFR enhancement/suppression occurs in clusters at $z\sim2$. One possible interpretation is that the SFR enhancement/suppression evolves in redshift. The star formation enhancement of normal star-forming galaxies has been reported in protoclusters at $z>2$ \citep{shimakawa18,ito20}, and the suppression in evolved clusters at $1<z<1.5$ \citep{old20}. Our cluster candidates at $z\sim2$ may be in the transition epoch of these two phases. Another explanation is the difference in halo masses. The expected halo masses of our clusters are at most $\log(M_{h}/M_{\odot})\sim14$ while those of \citet{old20}'s sample are $\log(M_{h}/M_{\odot})>14$. Since the quiescence of star formation strongly correlates with halo mass even at $z\sim1.5$ (\eg, \citealt{reeves21}), the lower halo mass of our sample may cause no detection of star formation suppression. 

Although there is no significant difference in the main sequence of the cluster and field galaxies, we still find an excess of quiescent galaxies in the cluster candidates with relatively high sSFR near the border (i.e. $\log{(\mathrm{sSFR}~\mathrm{[yr^{-1}]})}=-10$) compared to those in the field. 
These galaxies are possibly green valley galaxies, a population in transition from the star-forming phase to the quiescent phase (\eg, \citealt{schawinski14, noirot22}). Here we define green valley galaxies as those with $-10.5<\log{(\mathrm{sSFR~[yr^{-1}]})}<-10$ and calculate the fraction of galaxies that are in the green valley and above the mass limit ($\log{(M_*/M_\odot)} > 10.5$). 
We find the proportion of green valley galaxies in clusters is $(12.3\pm 2.8)$\%, much higher than that of field galaxies, which is $(2.01\pm 0.04)$\%. This trend qualitatively holds even when we slightly change the sSFR threshold for defining the green valley\footnote{For examples, when we apply a sSFR threshold of $-11.0 < \log{(\mathrm{sSFR~[yr^{-1}]})}<-10.5$ ($-11.0< \log{(\mathrm{sSFR~[yr^{-1}]})}<-10.0$), the quiescent fraction in the field is $(2.08\pm0.04)\%$, ($(4.10\pm0.06)\%$), while the quiescent fractions in the cluster environment is $(5.80\pm1.99)\%$ ($(18.1\pm3.28)\%$), respectively.}. 
This might be additional evidence that the star formation transition occurs more actively in the clusters. 
However, our data sets lack mid-infrared photometry or spectroscopy, such as $\mathrm{H}\alpha$ lines, which are essential for accurately characterizing their SFRs. We require more data to confirm this trend. 
We note \citet{mcnab21} have reported that, at $1<z<1.4$, there is no significant difference in the fractions of color-selected transition populations between GOGREEN clusters and the corresponding field at $\log{(M_*/M_\odot)> 10.5}$. Also, they have claimed that for low mass galaxies at $9.5<\log{(M_*/M_\odot)<10.5}$, there is a slight excess (5--10\%) of the transition population in the clusters compared to that in the field. Although a direct comparison is difficult because of different definitions of transition populations and stellar mass range, this might indicate the frequency or timescale of quenching changes with redshift. 

\subsection{Composite color-magnitude diagram \label{subsec:CCMD}}  
\begin{figure*}
\plotone{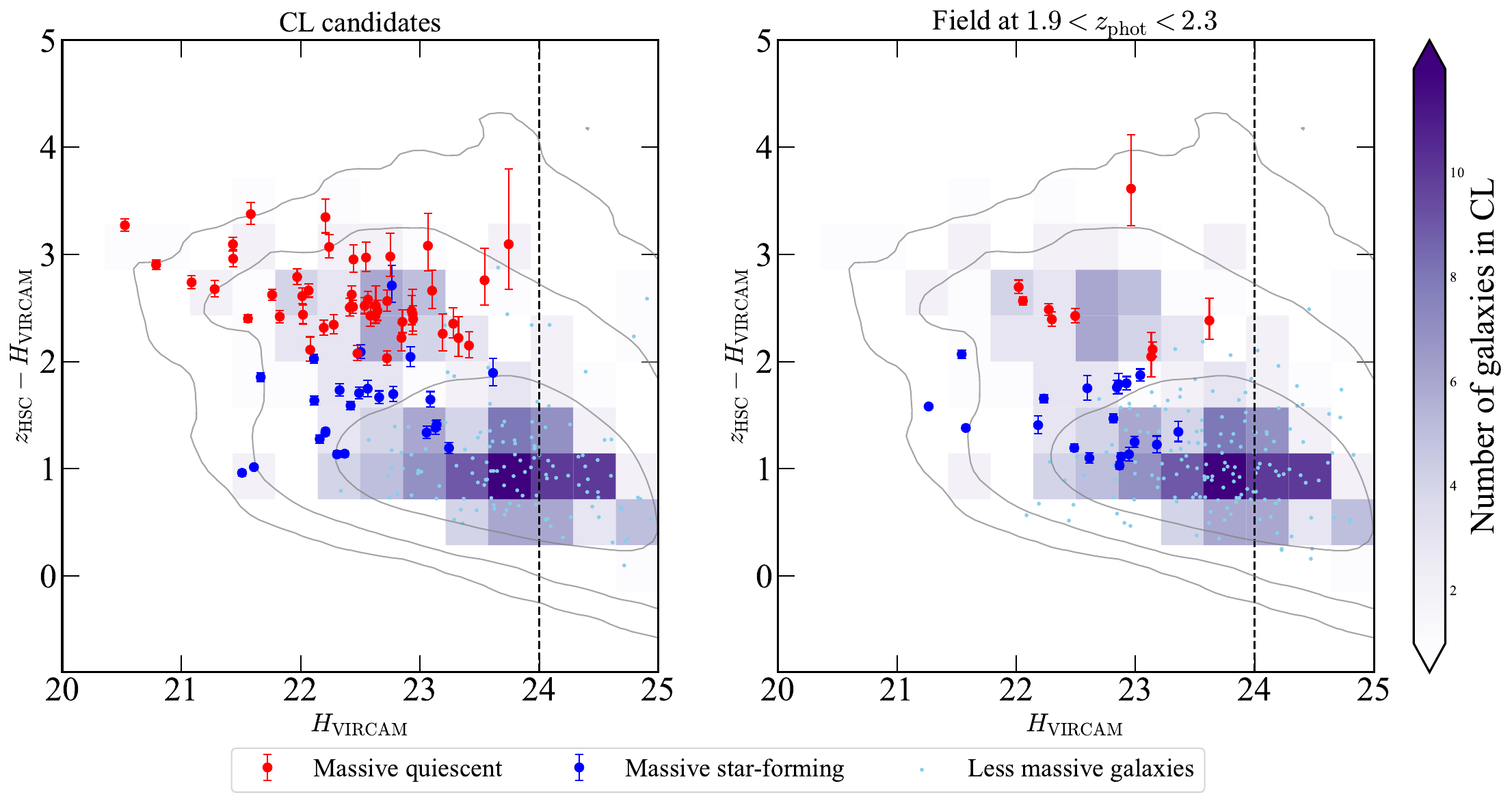}
\caption{
    The composite CMDs of the cluster candidates (left) and the field at $|z_\mathrm{phot} - 2.1| < 0.2$ (right), respectively. The red, blue, and light blue points are massive (${M_*} > 3 \times 10^{10} ~M_\odot$) quiescent, massive star-forming, and less massive galaxies, respectively. The quiescent fractions for massive galaxies are shown at the top of each panel. The grey contours show the distribution of all field galaxies at $z\sim2$ with 68\%, 95\%, and 99\%. The 2D histogram shows the number of field-subtracted galaxies in cluster candidates. Field subtraction is conducted in each grid of $z-H$ vs. $H$ plane for the 2D histogram. The white regions indicate that the number of galaxies after field subtraction is zero or below. 
\label{fig:CCMD}}
\end{figure*}

As shown in Figs.~\ref{fig:image1}--\ref{fig:image3}, the clear red sequence is seen in the CMD of each cluster candidate at $z\sim2$. Here we compare the composite CMD of clusters with that of the field to see the segregation of quiescent galaxies in the high redshift Universe. Fig.~\ref{fig:CCMD} shows the composite CMDs for cluster candidates (left) and field galaxies (right) at $z\sim2$. We show the cluster galaxies and the same number of randomly selected field galaxies. In addition, we also show smoothed field galaxy distributions by contours. 

Red sequences are seen in the CMDs of cluster candidates and the field. The dominance of massive quiescent galaxies (red dots) is much higher in the cluster environment, suggesting our selection successfully identifies systems dominated by quiescent galaxies at $z\sim2$. Moreover, there are very bright ($H_\mathrm{VIRCAM}\lesssim21.5$) massive quiescent galaxies in the cluster candidates, while there are a few such galaxies in the field. Conversely in the field, this bright end is dominated by massive star-forming galaxies. We note that this galaxy segregation is not always seen in general cluster samples at this redshift due to our sample being biased. 

\subsection{Quiescent fraction and quiescent fraction excess \label{subsec:QFE}} 
In the previous subsections, we show that the star-formation quenching is accelerated in our cluster environment. We further investigate the cause of the emergence of massive quiescent galaxies in high redshift clusters at $z\sim2$. 

\begin{figure*}
\gridline{\fig{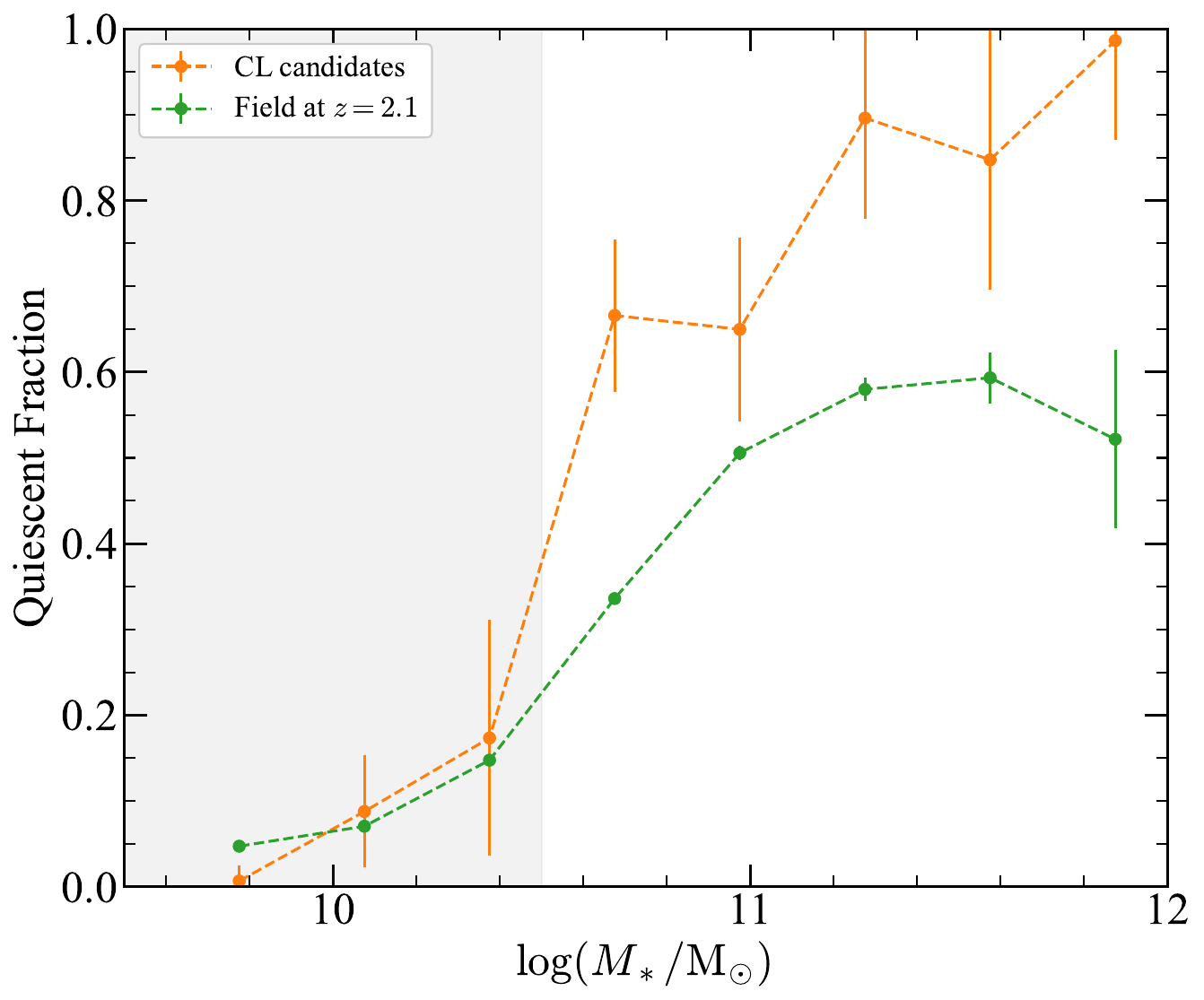}{0.48\textwidth}{}
          \fig{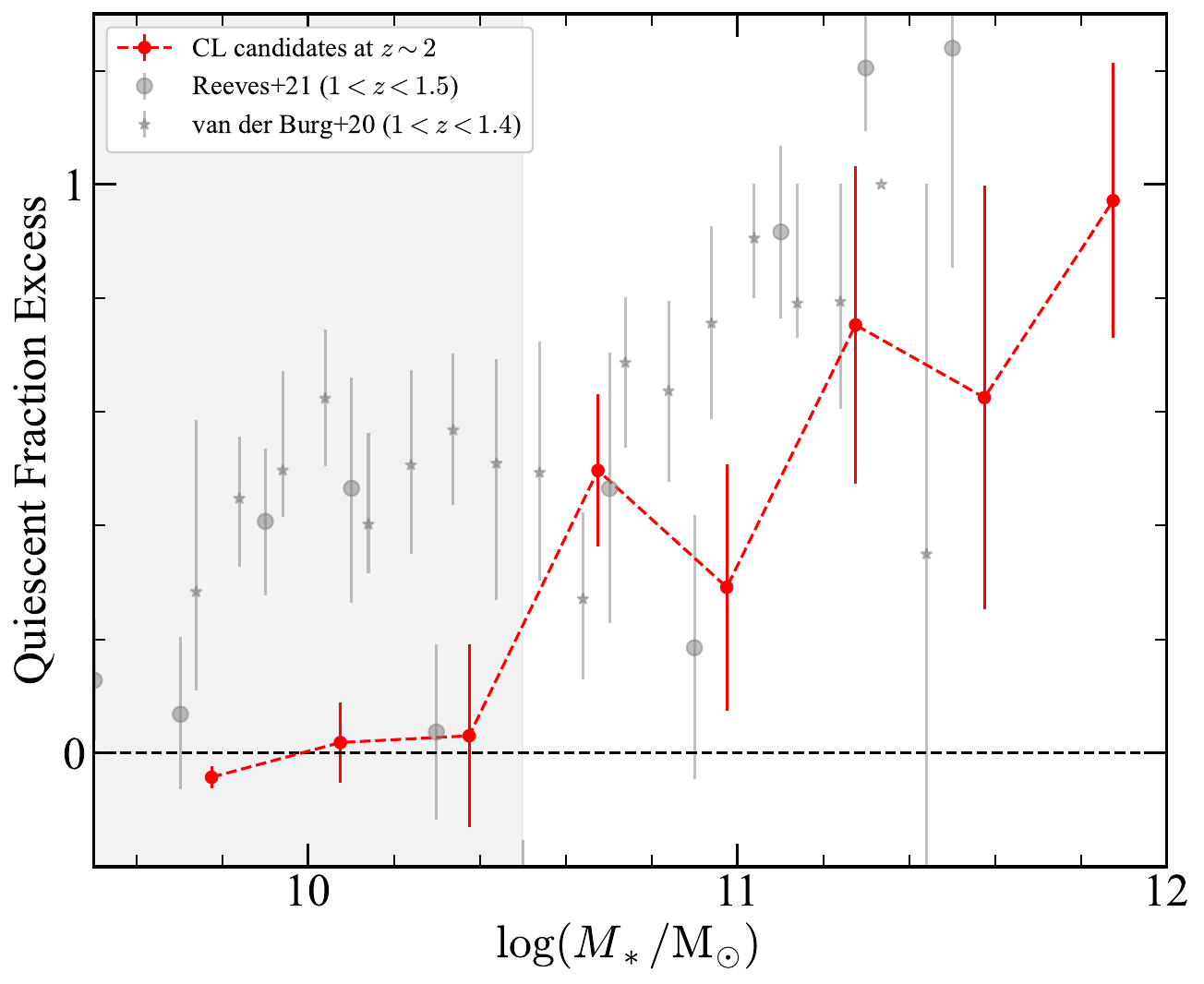}{0.48\textwidth}{}
          }
\caption{
\textit{Left}: The quiescent fraction against stellar mass. The orange and green points are the cluster and field galaxies, respectively. The grey shade indicates the mass range below the limiting mass. 
\textit{Right}: The quiescent fraction excess (QFE) against stellar mass. The red points show the QFE of our cluster candidates at $z\sim2$, while the gray symbols are for those of clusters/groups at $1<z<1.5$ from the literature \citep{vanderburg20,reeves21}.
\label{fig:QF}}
\end{figure*}

The left panel of Fig.~\ref{fig:QF} shows the quiescent fractions in the cluster candidates (orange) and the field at $z\sim2$ (green) as a function of stellar mass. 
The quiescent fraction $f_\mathrm{Q}$ is defined as:
\begin{equation}
    f_\mathrm{Q}(M_*) = \frac{N_\mathrm{Q}(M_*)}{N_\mathrm{Q}(M_*) + N_\mathrm{SF}(M_*)},
\end{equation}
where $N_\mathrm{Q}(M_*)$ and $N_\mathrm{SF}(M_*)$ are the number of quiescent galaxies and star-forming galaxies as a function of stellar mass $M_*$, respectively. 
As a general trend, more massive galaxies have higher quiescent fractions in both environments. In the whole mass range above the limiting stellar mass, the quiescent fraction of the cluster candidates is significantly higher than that of the field. 

To see the efficiency of environmental quenching, we additionally calculate the quantity called quiescent fraction excess (QFE)\footnote{Some terminologies in the literature represent the same quantity: conversion fraction (\eg, \citealt{balogh16}), environmental quenching efficiency (\eg, \citealt{nantais16}).} defined as: 
\begin{equation}
    \mathrm{QFE}(M_*) = \frac{f_\mathrm{Q,cluster}(M_*) - f_\mathrm{Q,field}(M_*)}{1 - f_\mathrm{Q,field}(M_*)}, 
\end{equation}
where $f_\mathrm{Q,cluster}(M_*)$ and $f_\mathrm{Q,field}(M_*)$ are the quiescent fraction of member galaxies in cluster candidates and field, respectively. 
QFE quantifies the excess fraction of the galaxies that are quenched in the cluster environment but would remain star-forming if they were in the field, compared to the star-forming galaxies in the field. 
We show the QFEs in the right panel of Fig.~\ref{fig:QF}. 
The QFE of the candidates increases with stellar mass, suggesting that more massive galaxies are more frequently quenched in the cluster environment at $z\sim2$. 
This trend is also reported at $z\lesssim1.5$ (\eg, \citealt{balogh16, vanderburg20, reeves21}). We confirm, for the first time, that this stellar-mass dependence of QFE holds in mature clusters up to $z\sim2$. 

For comparison, we overlay the QFEs derived by \citet{vanderburg20} and \citet{reeves21} in the right panel of Fig.~\ref{fig:QF}. They use the spectroscopically confirmed cluster/group samples at $1<z<1.5$. The QFE of our cluster candidates is systematically lower than that of \citet{vanderburg20}, which has used a more massive ($\log{(M_{200}/M_\odot)}>14$) and low redshift cluster sample than ours. Meanwhile, our cluster candidates have a similar QFE value to \citet{reeves21}. The halo masses of \citet{reeves21}'s group sample are $13.6 < \log{(M_\mathrm{200}/M_\odot)}< 14.0$, comparable to the X-ray inferred mass of the CL1 ($\log{(M_{200}/M_\odot)} = 13.9$). This might indicate the similarity of the evolutionary phase between our cluster candidates and those of \citet{reeves21}. 

\citet{papovich18} have suggested the stellar-mass dependent quenching based on the stellar mass function at $1.5<z<2$ in a top quartile high-density environment. However, \citet{edward24} have not found evidence of mass-dependent quenching in their protocluster samples at $2.0<z<2.5$. At even higher redshifts ($z>2$), several individual protocluster studies also indicate the presence of mass-dependent quenching (e.g., \citealt{mcconachie22}). 
Our findings align with these previous studies. Compared to them, we construct a statistical sample of seven cluster candidates to capture the highest density environment at $z\sim2$ and provide statistical evidence of mass-dependent quenching.

The physical origins of the excess quenching shown above may provide a clue to understanding the dominant quenching scenario in high-redshift clusters. Here, we briefly discuss a few possible quenching paths. In local massive clusters, cold and/or hot gas removal from galaxies due to ram-pressure stripping (RPS) is a significant source of quenching (see \citealt{boselli22} for a comprehensive review). 
Since we detect the extended X-ray emission from CL1, RPS may work at least in this candidate. 
Theoretical predictions suggest that the density of ICM can increase with redshift for a fixed cluster mass \citep{fujita01,boselli22}. The high ICM density can lead to stronger RPS in a high-redshift cluster at a given halo mass. 
On the other hand, the stellar components of galaxies are more compact in the higher redshift Universe \citep{vanderWel14}. This implies that the associated cold gas may be more strongly gravitationally bound and harder to strip than lower redshift galaxies. 
Since the balance between these two is not clear, we cannot conclude if RPS effectively works in our clusters. 

Galaxy-galaxy interaction is also thought to be a cause of quenching, which may trigger star-burst or AGN activity (\eg, \citealp{moore96,moore98,man18}). If we carefully check the morphologies of member galaxies of the candidate clusters in future work, we might find disturbed structures or tidal tails typical for merging systems. If we derive the detailed star formation history of member galaxies inferred from spectroscopic data, we may check whether intense star formation frequently occurs in clusters. Spectra can also provide AGN signatures, which will help us constrain the contribution of AGNs for quenching. 

Lastly, we attempt to explain the observed trends in quiescent fraction and QFE in the `overconsumption' scenario (\citealt{mcgee14, balogh16}). 

In a halo-mass system with $\log{(M_\mathrm{halo}/M_\odot)} \gtrsim 12$, cold gas that is accreting onto the halo gets shock-heated to the virial temperature (e.g., \citealt{dekel06}). When a satellite galaxy enters a large hot halo of a cluster, a gas supply to the satellite galaxy can be cut off. Under this condition, galaxies quench their star formation shortly after their gas depletion time (i.e., gas mass divided by SFR). \citet{mcgee14} showed that star formation-driven winds in satellite galaxies can lead to a short quenching time without cosmological accretion to cluster halos. 
Therefore, cluster galaxies can rapidly consume their gas and quench earlier than the field. 
This condition gives a higher quiescent fraction of cluster galaxies than that of galaxies in fields. 
\citet{balogh16} have shown that the delay times (the time to starve a galaxy’s gas reservoir) can be shorter in high-mass galaxies than low-mass ones under the assumption that their gas is ejected from their halo with a strength proportional to their SFR. 
The shorter delay times for higher-mass galaxies result in preferential quenching of massive galaxies under the no-gas-supply condition (i.e., the higher QFE for higher-mass galaxies). 

Quenching paths discussed above are not confirmed yet in this study because only photometric data is currently available to infer galaxy properties. We need to carry over follow-up observations to examine possible quenching paths from confirmed cluster members, the star formation history, and the presence or absence of AGNs. 

\section{Summary and conclusions} \label{sec:conclusions}

We conduct a survey targeting galaxy clusters at the frontier redshift of $z\sim2$ in the $\sim 3.5\,\mathrm{deg}^{2}$ area of the XMM-LSS field. Applying a Bayesian photometric redshift code \citep{tanaka15} to the $u^*$-band through the $K$-band photometry gathered from the HSC-SSP, CLAUDS, and VIDEO surveys, we estimate photometric redshifts of galaxies and their physical properties, such as stellar mass and SFR. We then adopt a Gaussian kernel density estimate with two kernel widths, $10\arcsec$ and $60\arcsec$, to draw a density map of massive quiescent galaxies. Based on this density map, we search for significant overdensities with red sequences. Our discoveries are as follows:  
\begin{enumerate}
    \item We identify seven prominent overdensities of massive quiescent galaxies as cluster candidates at $z \sim 2$. These candidates reside in density peaks of massive quiescent galaxies with $\sigma_{60\arcsec} \gtrsim 6$ and $\sigma_{10\arcsec} \gtrsim 100$. We confirm there are high spatial and redshift concentrations of the member galaxies within a radius of 0.5 physical Mpc and $\Delta z=\pm0.2$, respectively. Additionally, we find clear red sequences of member galaxies in color-magnitude diagrams ($z-H$ vs. $H$), suggesting our cluster candidates are evolved systems. 
    \item One of the candidates (CL1) shows an extended X-ray emission, indicating its well-evolved and virialized nature. The X-ray luminosity is $(1.46\pm0.35) \times 10^{44}$ erg s$^{-1}$ in $0.1$--$2.4\,\mathrm{keV}$, with the virial mass of $M_{200} = (7.75\pm1.15) \times 10^{13} ~M_\odot$. An X-ray point source is detected near the center of CL2 (cf. XMM05036 of the XMM-SERVS point source catalog; \citealt{chen18}), suggesting that the BCG is an AGN. 
    \item The stellar mass distribution of the cluster galaxies is skewed towards higher masses compared to those of field galaxies, suggesting accelerated galaxy growth in the cluster regions. We examine the SFR-$M_{*}$ relation for star-forming galaxies (i.e., the star formation main sequence) and find there is no significant difference between the cluster and the field. We also check the fraction of galaxies with $-10.5<\log{(\mathrm{sSFR}~\mathrm{[yr^{-1}]})} < -10$ that may be during the quenching process (i.e., green valley galaxies). Then we find that a larger fraction ($\sim$12\%) of galaxies is located at this sSFR range in the cluster candidates than in the field ($\sim$2\%). This indicates that star formation transition is actively proceeding in some evolved clusters even at $z\sim2$. 
    \item The quiescent fraction of massive ($\log(M_{*}/M_{\odot})>10.5$) galaxies in our cluster candidates is higher than that of the field, indicating the galactic `segregation' in the early Universe even at $z\sim2$. 
    Moreover, the QFE of these candidates shows an increasing trend with stellar mass. 
    This is the first direct evidence with the statistical cluster samples that show mass-dependent environmental quenching in massive clusters at $z\sim2$. 
    This trend can be explained qualitatively by the `overconsumption' scenario. However, we need follow-up observations to discuss the quenching scenario in detail with inferred star formation history or star formation efficiency. 
\end{enumerate}

Cluster candidates at $z\sim2$ identified in this work will provide key information to understanding galaxy evolution in the densest environments in the early Universe. We plan to expand our cluster/group survey to other deep fields in the multiwavelength survey of the HSC-SSP, making massive cluster samples at $z\sim2$ roughly five times larger. This work is based on the $u^*$-band to $K_\mathrm{s}$-band photometry. If we add mid-IR to far-IR photometry from Spitzer/IRAC and Multiband Imaging Photometer for the Spitzer (MIPS; \citealt{rieke04}), we may be able to target clusters/groups at even higher redshifts. It also enables us to better characterize the stellar population and dust properties. Spectroscopic follow-up is important not only to confirm cluster memberships but also to infer the formation history of each cluster galaxy. At this point, our candidates are good targets for the upcoming Subaru/Prime Focus Spectrograph (PFS; \citealt{tamura16}) surveys. In addition, the Atacama Large Millimeter/submillimeter Array (ALMA) is also a powerful instrument that helps characterize dust and molecular gas content, allowing for the estimation of the star formation efficiency. Our cluster candidates at $z\sim2$ will trigger statistical analyses of the environmental quenching in overdense regions at $z\gtrsim2$. 

\section{Acknowledgments} \label{ack}

We appreciate the anonymous referee for the useful comments that improved our manuscript.

This work is partly carried out as a part of the Summer Student Program held by the Department of Astronomical Science, The Graduate University for Advanced Studies, SOKENDAI, and the National Astronomical Observatory of Japan (NAOJ) in 2022. We thank the SOKENDAI Astronomical Science Program for their support. 

MA acknowledges that this work was supported by the Data-Scientist-Type Researcher Training Project of The Graduate University for Advanced Studies, SOKENDAI.

The Hyper Suprime-Cam (HSC) collaboration includes the astronomical communities of Japan and Taiwan, and Princeton University. The HSC instrumentation and software were developed by the National Astronomical Observatory of Japan (NAOJ), the Kavli Institute for the Physics and Mathematics of the Universe (Kavli IPMU), the University of Tokyo, the High Energy Accelerator Research Organization (KEK), the Academia Sinica Institute for Astronomy and Astrophysics in Taiwan (ASIAA), and Princeton University. Funding was contributed by the FIRST program from the Japanese Cabinet Office, the Ministry of Education, Culture, Sports, Science and Technology (MEXT), the Japan Society for the Promotion of Science (JSPS), Japan Science and Technology Agency (JST), the Toray Science Foundation, NAOJ, Kavli IPMU, KEK, ASIAA, and Princeton University. 

This paper makes use of software developed for Vera C. Rubin Observatory. We thank the Rubin Observatory for making their code available as free software at http://pipelines.lsst.io/.

This paper is based on data collected at the Subaru Telescope and retrieved from the HSC data archive system, which is operated by the Subaru Telescope and Astronomy Data Center (ADC) at NAOJ. Data analysis was in part carried out with the cooperation of Center for Computational Astrophysics (CfCA), NAOJ. We are honored and grateful for the opportunity of observing the Universe from Maunakea, which has the cultural, historical and natural significance in Hawaii. 

This work is also based on data products from observations made with ESO Telescopes at the La Silla Paranal Observatory as part of the VISTA Deep Extragalactic Observations (VIDEO) survey, under programme ID 179.A-2006 (PI: Jarvis). 

These data were obtained and processed as part of the CFHT Large Area U-band Deep Survey (CLAUDS), which is a collaboration between astronomers from Canada, France, and China described in \citet{sawicki19}. CLAUDS data products can be accessed from https://www.clauds.net. CLAUDS is based on observations obtained with MegaPrime/ MegaCam, a joint project of CFHT and CEA/DAPNIA, at the CFHT which is operated by the National Research Council (NRC) of Canada, the Institut National des Science de l’Univers of the Centre National de la Recherche Scientifique (CNRS) of France, and the University of Hawaii. CLAUDS uses data obtained in part through the Telescope Access Program (TAP), which has been funded by the National Astronomical Observatories, Chinese Academy of Sciences, and the Special Fund for Astronomy from the Ministry of Finance of China. CLAUDS uses data products from TERAPIX and the Canadian Astronomy Data Centre (CADC) and was carried out using resources from Compute Canada and Canadian Advanced Network For Astrophysical Research (CANFAR).

\vspace{5mm}

\software{Astropy \citep{astropy:2013, astropy:2018, astropy:2022},  
          Matplotlib \citep{hunter07},
          NumPy \citep{harris20},
          SciPy \citep{virtanen20},
          pandas \citep{mckinney2010data}
          }

\bibliography{reference}{}
\bibliographystyle{aasjournal}

\end{document}